\documentclass[pre,floatfix,superscriptaddress,twocolumn,showpacs]{revtex4-1}
\usepackage{graphicx}

\usepackage[utf8]{inputenc}
\usepackage[T1]{fontenc}
\usepackage{graphicx}
\usepackage{amsmath,amssymb}
\usepackage{hyperref}
\usepackage{nameref}
\usepackage{xcolor}
\hypersetup{
    colorlinks=true,
    linkcolor=blue,
    citecolor=blue,
    urlcolor=blue
}
\usepackage{tensor}
\usepackage{textgreek}
\usepackage{bm}
\usepackage{mathrsfs}

\usepackage{xcolor} 

\usepackage{multibib}
\newcites{S}{Supplemental References}

\makeatletter
\renewcommand{\@fnsymbol}[1]{%
  \ifcase#1\or * \or †\or ‡\or **\or \ddagger\or \mathsection\or \mathparagraph\or \|\or **\else\@arabic{#1}\fi}
\makeatother

\begin{document}
\title{Persistence-Driven Void Formation in Dense Active–Passive Mixtures}
\author{Giulia Janzen}
\email{gjanzen@ucm.es}
\affiliation{Department of Theoretical Physics, Complutense University of Madrid, Madrid, 28040, Spain}
\author{Liesbeth M.~C.~Janssen}
\affiliation{Department of Applied Physics, Eindhoven University of Technology, P.O.~Box 513, 5600 MB Eindhoven, The Netherlands}
\author{Nuno A. M. Ara\'ujo}
\affiliation{Centro de Física Teórica e Computacional, Faculdade de Ciências, Universidade de Lisboa, 1749-016 Lisboa, Portugal}
\affiliation{Departamento de F\'sica, Faculdade de Ciências, Universidade de Lisboa, 1749-016 Lisboa, Portugal}
\author{Rastko Sknepnek}
\affiliation{Faculty of Life Sciences, University of Dundee, Dundee, DD1 5EH, United Kingdom}
\affiliation{Faculty of Science, Engineering and Business, University of Dundee, Dundee, DD1 4HN, United Kingdom}
\author{D. A. Matoz-Fernandez}
\email{dmatoz@ucm.es}
\affiliation{Department of Theoretical Physics, Complutense University of Madrid, Madrid, 28040, Spain}

\begin{abstract}
It is well established that dilute active dopants can melt an arrested amorphous solid by enhancing cage breaking and accelerating structural relaxation. Yet it remains unclear whether increasing persistence simply amplifies this effective melting or instead reorganizes the fluidization mechanism itself. Here we show that, in a minimal active–passive mixture, increasing persistence drives a crossover from homogeneous fluidization to a localized mechanical instability, demonstrating that sustained active forcing restructures relaxation in space rather than merely strengthening it. Persistent dopants accumulate stress and nucleate voids as their mechanically perturbed regions overlap. In this regime, rearrangements localize at void boundaries, and active and passive particles exhibit comparable mobility, producing dynamics reminiscent of crowd mosh pits. Persistence therefore reorganizes fluidization through stress accumulation and confinement, revealing a distinct nonequilibrium localization mechanism in disordered solids.
\end{abstract}
\maketitle
Mixtures of particles with different motilities offer a minimal setting to probe how nonequilibrium driving reshapes collective behavior in active matter~\cite{marchetti2013hydrodynamics}. Such systems arise across physical and biological contexts, including driven colloids, granular media, bacterial suspensions, and living tissues, where constituents with heterogeneous dynamics interact in dense, disordered environments~\cite{grosser21,coban21,dinelli2023non,grober2023unconventional}. How activity couples to structural relaxation, mechanical rigidity, and self-organization near the glass transition remains a central challenge in nonequilibrium statistical physics~\cite{RevModPhys.88.045006,berthier2019glassy,janssen2019active}.

Motility heterogeneity has been explored extensively, ranging from tracer dynamics in active or passive baths, where anomalous transport has been reported~\cite{PhysRevLett.103.198103,C0SM00164C,PhysRevE.88.041002,mejia2011bias,benichou2013biased,Wittkowski_2017},
to dense active-passive mixtures exhibiting cooperative motion and activity-induced structural rearrangements~\cite{vasilyev2017cooperative,Banerjee2022,Lozano2019,PhysRevLett.122.148101,D1SM01092A,janzen23}.
In dense regimes, such mixtures exhibit crystallization in hard-sphere glasses~\cite{Ni2014_Crystallizing}, segregation of anisotropic particles~\cite{C2SM06960A}, activity-induced phase separation~\cite{PhysRevLett.112.198301,PhysRevLett.114.018301}, and collective states including flocking and mesoscale turbulence~\cite{C4SM01562B}.
More recently, attention has turned to the dilute limit, where a small fraction of active dopants modifies the mechanical response and relaxation pathways of an otherwise passive amorphous host~\cite{Paoluzzi2024,omar2018swimming}.

Activity in dense amorphous media is commonly interpreted as an effective temperature that facilitates structural relaxation and drives glass-fluid transitions~\cite{mandal2016active,MatozPRL2017,Shee2025dopants}. However, this mapping is not universal, breaking down when active forces are sufficiently persistent, a regime in which the injected stress is not isotropic noise but a structured, long-lived perturbation of the mechanical environment~\cite{levis2015single,PetrellieffectiveT2020,mandal2020multiple,JanzenAging2022,PhysRevLett.131.178302}. This mechanical picture is supported by studies of driven intruders, where strongly forced particles generate persistent structural defects in the surrounding medium rather than merely amplifying thermal fluctuations~\cite{Reichhardt2004PhysRevLett}. 

More broadly, dense active matter shares deep connections with athermally driven disordered solids, with persistent active forces acting as perturbations of the underlying energy landscape rather than isotropic noise~\cite{morse2021direct}. Consistent with this view, active-passive and hot-cold mixtures exhibit effective interactions, demixing, and fluidization phenomena that cannot be reproduced by uniformly raising the temperature~\cite{PhysRevFluids.2.043103,PhysRevE.92.032118,FreyPRL2016,ManningPRL2024}. These effects become particularly pronounced when the persistence time of the active forcing exceeds the elastic relaxation time of the host medium~\cite{Matoz2020,Shee2025dopants}. In this regime, injected stresses do not fully relax between reorientation events, allowing stress to accumulate and spread over mesoscopic distances. Whether persistence simply enhances structural relaxation or instead reorganizes the mechanical response of the host medium remains an open question.

In this Letter, we show that sufficiently persistent, dilute active dopants trigger a persistence-controlled mechanical instability in dense amorphous solids. Increasing persistence drives a crossover from a homogeneous fluid with spatially uniform rearrangements to a void-forming state in which accumulated stress nucleates low-density regions and reorganizes transport along their boundaries. In this regime, active and passive particles attain comparable mobility, reminiscent of mosh-pit crowd dynamics~\cite{PhysRevLett.110.228701}. Moreover, passive particles undergo intermittent particle-scale rotations without global chiral order. This emergent chirality arises from confinement by self-generated voids and is distinct from fluidization induced by intrinsically chiral active particles~\cite{Debetschirality2023}.

We studied a two-dimensional dense amorphous solid of $N=10^4$ polydisperse disks at packing fraction $\phi=0.9$, i.e., deep in the dynamically arrested regime. To avoid crystallization, radii $a_i$ are drawn from a uniform distribution with mean $a$ and half-width $0.2\sqrt{3}a$, corresponding to $20\%$ polydispersity. Particles interact via purely repulsive harmonic forces, $\mathbf{F}_{ij}=k(a_i+a_j-r_{ij})\hat{\mathbf r}_{ij}$ for $r_{ij}<a_i+a_j$, and $\mathbf{F}_{ij}=\mathbf{0}$, otherwise. Starting from a quenched amorphous configuration, we randomly select $N_a$ particles and convert them into active Brownian particles (ABPs), corresponding to active fractions $\phi_a=N_a/N$. Active particles self-propel with speed $v_0$ along $\mathbf n_i=(\cos\theta_i,\sin\theta_i)$, where $\theta_i$ is the angle with the $x-$axis of the lab frame, while their orientations undergo rotational diffusion, $\dot\theta_i=\sqrt{2D_r}\,\eta_i$, where $D_r$ is the rotational diffusion coefficient and $\eta_i$ Gaussian white noise. This defines the persistence time $\tau_r=D_r^{-1}$ and persistence length $l_p=v_0\tau_r$. Passive particles have $v_0=0$ and are thermalized by translational Brownian noise.

Lengths are measured in units of $a$ and times in units of the elastic relaxation time $(\mu k)^{-1}$, where $\mu$ is the mobility. The corresponding dimensionless control parameters are $\tilde v=v_0/(a\mu k)$ and $\tilde D_r=D_r/(\mu k)$. We fix $\tilde v=0.1$ and tune the persistence length $l_p$ by varying $\tilde D_r\in[5\times10^{-5},\,5\times10^{-1}]$, and integrate the overdamped dynamics with time step $\mathrm{d}t=5\times10^{-3}$ using the GPU-accelerated \textsc{SAMoS} molecular dynamics package~\cite{SAMoS2024}. Unless stated otherwise, passive particles have translational diffusivity $\tilde D_t=10^{-7}$ and ensemble averages $\langle\cdots\rangle$ are computed over $10^2$ realizations. We choose parameters such that a fully active system at the same $\phi$ lies above the glass transition throughout this range~\cite{C3SM52469H}.

\begin{figure}[t]
\includegraphics[width=1\columnwidth]{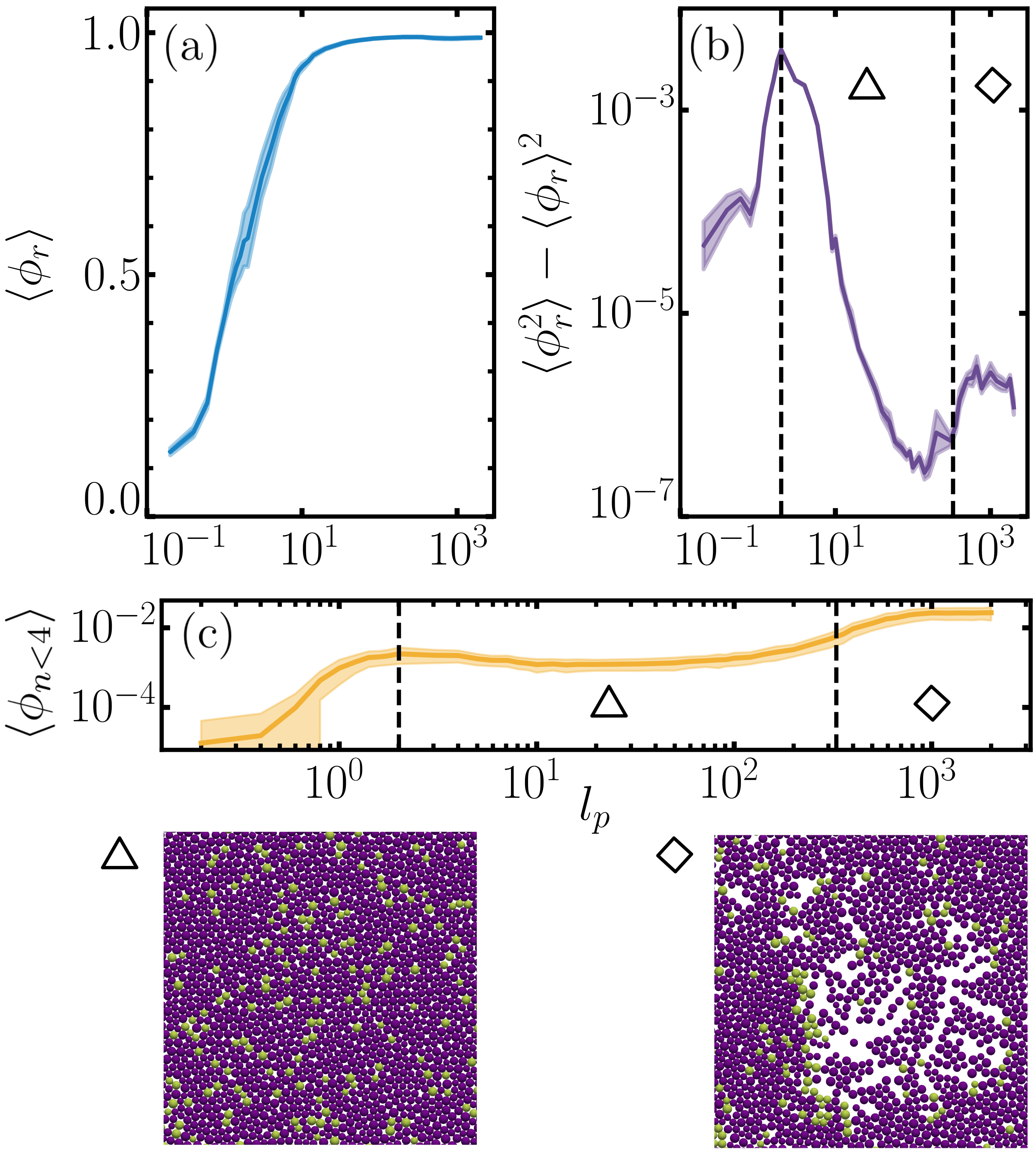}
\caption{Characterization of the system at fixed active fraction $\phi_a = 10^{-1}$ as a function of the persistence length $l_p$.
(a) Mean rearrangement fraction $\langle \phi_r \rangle$.
(b) Rearrangement fluctuations $\chi_r = \langle \phi_r^2 \rangle - \langle \phi_r \rangle^2$. The first peak marks the arrest–fluidization crossover; at larger $l_p$, a second peak signals the onset of a dynamically distinct void-forming regime.
(c) Mean fraction of undercoordinated particles $\langle \phi_{n<4} \rangle$. A plateau between the two peaks is followed by renewed growth near the second peak, consistent with void nucleation. Shaded bands indicate the standard deviation; solid lines are guides to the eye. Triangle and diamond symbols mark the homogeneous-fluid and void-forming regimes, respectively, and correspond to the representative snapshots. In the snapshots, passive particles are shown in purple and active particles in yellow.}
\label{fig:1}
\end{figure}

Figure~\ref{fig:1} characterizes the system dynamics as a function of the persistence length $l_p$ for a representative case $\phi_a=10^{-1}$. In Fig.~\ref{fig:1}(a), we show the rearrangement fraction $\langle \phi_r \rangle$, defined as the fraction of particles whose nearest-neighbor list changes between times $t$ and $t+\Delta t$, with $\Delta t = 2\times 10^{3}$. For larger $l_p$, the overall level of rearrangements increases, and by $l_p\sim 10^{1}$ the system fluidizes.
To probe the character of this fluidized state, we examine the fluctuations $\chi_r\equiv\langle \phi_r^2\rangle-\langle \phi_r\rangle^2$ [Fig.~\ref{fig:1}(b)]. Rather than growing monotonically, $\chi_r$ exhibits two distinct features: a first peak at the onset of fluidization, followed by a decrease at intermediate persistence and a renewed increase at large persistence, $l_p\sim10^{3}$. The second increase signals that the strongly persistent regime is dynamically distinct from the intermediate-persistence homogeneous fluid.

To understand the structural origin of the non-monotonic behavior of $\chi_r$, we analyze the distribution of the local coordination number $n$, defined as the number of nearest neighbors of a particle (see Supplemental Material~\cite{SI}). At large $l_p$, the nearest-neighbor distribution develops enhanced weight at low coordination (small $n$), indicating a growing population of undercoordinated particles that preferentially localize at boundaries of low-density regions, a direct signature of void-like structures. Consistently, $\langle\phi_{n<4}\rangle$, the fraction of particles with fewer than four nearest neighbors [Fig.~\ref{fig:1}(c)], increases again at large $l_p$,  mirroring the non-monotonic behavior of $\chi_r$. Direct measurement of the largest void area~\cite{SI} confirms that this regime corresponds to void nucleation.
Together, these results reveal two persistence-controlled crossovers. At intermediate persistence, the arrested solid first fluidizes, whereas at stronger persistence, the system transitions to a void-forming state.
\begin{figure}[t]
    \centering
\includegraphics[width=\columnwidth]{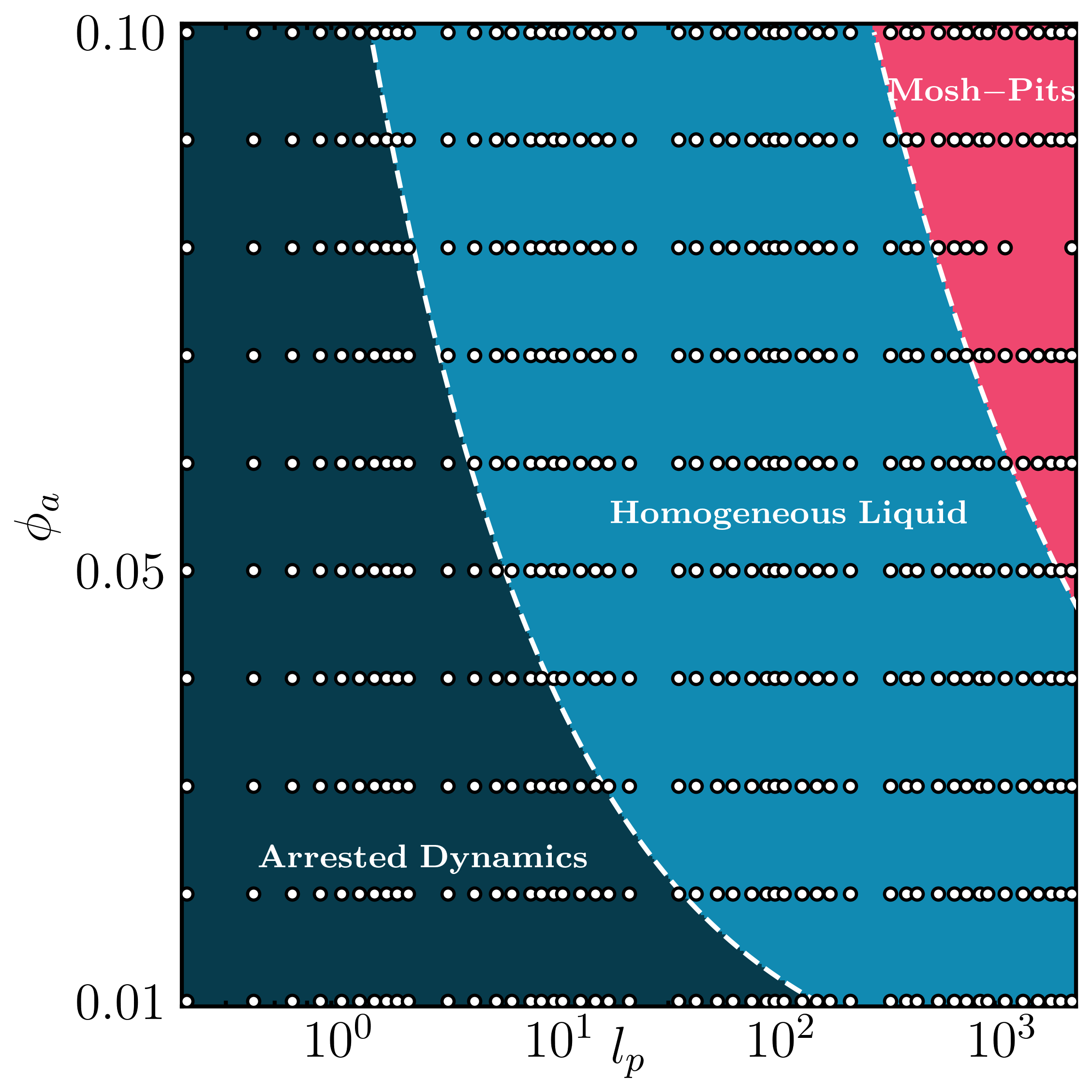}
\caption{Phase diagram of the active–passive mixture in the $(l_p,\phi_a)$ plane. Dark blue denotes arrested dynamics, light blue a homogeneous liquid, and purple the void-forming (mosh-pit-like) regime. Boundaries are determined from the rearrangement fluctuations $\chi_r = \langle \phi_r^2 \rangle - \langle \phi_r \rangle^2$. The arrest–fluidization boundary corresponds to the first peak in $\chi_r$, and the homogeneous–void-forming boundary to the second peak. White dashed lines show power-law fits of the form $l_p = C \phi_a^{b}$ to the extracted peak locations. Symbols denote simulation points.}
    \label{fig:2}
\end{figure}

Having identified two persistence-controlled crossovers, we analyze how these regimes evolve as a function of the dopant fraction $\phi_a$. In Fig.~\ref{fig:2}, we summarize the behavior in the $(l_p,\phi_a)$  plane. We extract regime boundaries from $\chi_r$, identifying the arrest–fluidization line with the first maximum of $\chi_r$, and the onset of void formation with its renewed increase at large persistence. Three regimes emerge. At small persistence, the system remains arrested within our observation window, i.e., $\Delta t = 2\times 10^{3}$. At intermediate persistence, it forms a homogeneous fluid consistent with an effective-temperature description, occupying a broad region of the diagram. At large persistence and sufficient $\phi_a$, a second crossover marks the void-forming regime, in which highly persistent active forcing nucleates self-generated low-density regions, reminiscent of crowd mosh pits. 

Strikingly, the boundary separating the homogeneous and void-forming regimes follows a robust inverse scaling $l_p^\star \sim \phi_a^{-1}$ (see Supplemental Material~\cite{SI}). This scaling reflects the mechanical origin of the transition. In overdamped amorphous media, localized forcing generates stress and displacement fields that propagate through the material and relax on finite timescales, as captured by stress-diffusion and elastoplastic models~\cite{Nicolas2018,MatozPRL2017}. Recent work has shown that even a dilute population of active dopants can induce continuum-scale mechanical correlations in dense athermal hosts, producing long-range velocity correlations with a persistence-controlled correlation length~\cite{abbaspour2024long}. Similarly, each dopant injects stress into the passive medium over a persistence time, creating a mechanically perturbed region whose size grows with $l_p$ until stress relaxation dissipates it. This suggests that void nucleation is a collective mechanical instability rather than a geometric percolation process. 

\begin{figure}[h!]
\includegraphics[width=1\columnwidth]{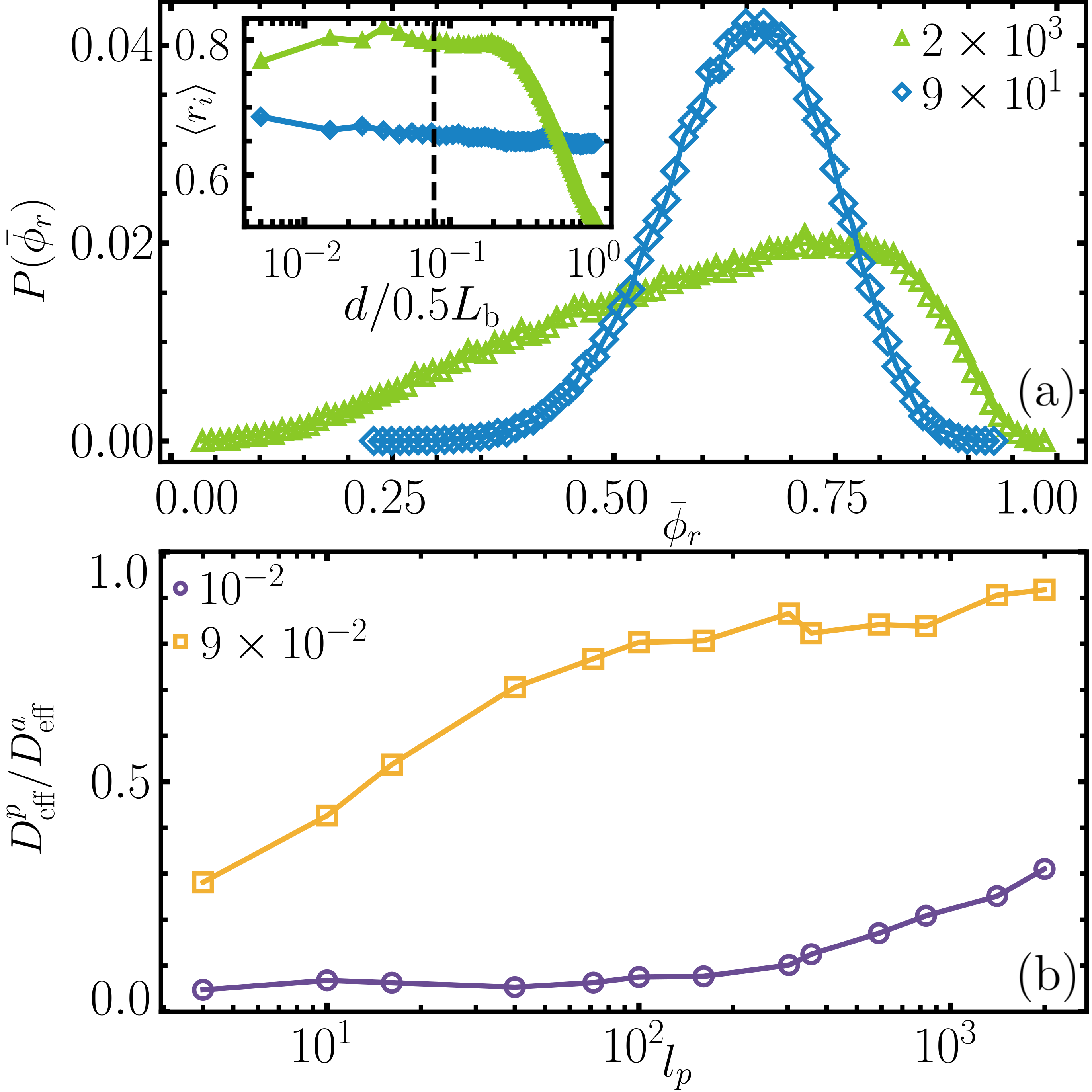}
\caption{Characterization of the two fluid regimes at $\phi_a = 10^{-1}$. Passive particles are shown in purple, active particles in yellow.
(a) Probability distribution of the coarse-grained rearrangement fraction $P(\bar{\phi}_r)$ for $l_p = 9\times 10^{1}$ (green triangles; homogeneous fluid) and $l_p = 2\times 10^{3}$ (blue diamonds; void-forming regime), computed over a time window $\Delta t = 10^{2}$. Inset: mean per-particle rearrangement rate $\langle r_i \rangle$ as a function of distance from the void center (in units of $L_b/2$, where $L_b$ is the box size). The dashed line marks the void radius; $\langle r_i \rangle$ peaks sharply at the void boundary and is suppressed in the bulk. In the homogeneous fluid, the same analysis referenced to a random point yields an approximately flat profile.
(b) Ratio of effective diffusion coefficients $D_{\mathrm{eff}}^{p}/D_{\mathrm{eff}}^{a}$ versus $l_p$ at $\phi_a = 10^{-2}$ (purple circles) and $\phi_a = 9\times 10^{-2}$ (yellow squares).}
\label{fig:3} 
\end{figure}

To rationalize the measured boundary, we construct a minimal scaling argument based on stress diffusion and overlap of dopant-induced perturbation regions. In the dilute limit, a single persistent dopant injects stress over one persistence time but remains
below the threshold for irreversible rearrangements. Using a minimal overdamped Maxwell viscoelastic closure and
eliminating the velocity field yields a diffusion-relaxation equation for the coarse-grained stress, with stress
diffusivity $D_\sigma = G/\zeta$ and relaxation time $\tau_m$ (see Supplemental Material~\cite{SI}). For persistence
times not exceeding relaxation ($\tau_r \lesssim \tau_m$), stress spreads diffusively over a radius
$r_{\rm infl} \sim \sqrt{D_\sigma \tau_r}$, defining a mechanical influence area
$V_{\rm infl} \sim D_\sigma \tau_r$~\cite{morse2021direct,Agoritsas2024memory}.

Void onset occurs when these mechanically perturbed regions overlap to form a system-spanning mechanically correlated region, i.e., when $\rho_a V_{\rm infl} \sim \mathcal{O}(1)$, where $\rho_a$ is the number density of active particles. Using
$l_p = v_0 \tau_r$ and $\rho_a \propto \phi_a$, this yields the scaling $l_p^\star \sim \phi_a^{-1}$, in agreement
with simulations (Fig.~S1 in the Supplemental Material~\cite{SI}). The observed inverse scaling rules out ballistic geometric overlap
($l_p^\star \sim \phi_a^{-1/2}$) and identifies diffusive stress accumulation and collective reinforcement as the
operative mechanism driving void formation.

We next identify the microscopic distinction between the two fluid states. 
In Fig.~\ref{fig:3}(a), we compare a representative homogeneous fluid ($l_p=9\times10^{1}$) with a void-forming state ($l_p=2\times10^{3}$) at fixed $\phi_a=10^{-1}$. The distribution of coarse-grained rearrangement activity $P(\bar{\phi}_r)$ (where $\bar{\phi}_r$ is the rearrangement fraction averaged over spatial bins within a time window $\Delta t$) reveals a qualitative change. The homogeneous fluid exhibits a narrow, nearly symmetric distribution consistent with spatially uniform dynamics; the void-forming regime develops a broad, strongly skewed distribution, indicating highly heterogeneous rearrangements. This heterogeneity has a spatial structure. The mean per-particle rearrangement rate as a function of distance from the void center (see inset, Fig.~\ref{fig:3}(a); see~\cite{SI} for more details on void identification) peaks sharply at the void boundary and is suppressed in the bulk. By contrast, the homogeneous fluid yields an approximately flat profile referenced to a random point, consistent with spatially uniform rearrangements.

Rearrangement localization reorganizes transport. In Fig.~\ref{fig:3}(b), the ratio of effective diffusion coefficients $D_{\mathrm{eff}}^{p}/D_{\mathrm{eff}}^{a}$ remains below unity in the homogeneous fluid, indicating that active particles are more mobile. In contrast, in the void-forming regime, the ratio approaches unity, as persistent dopants drag neighboring passive particles along the void boundary, equalizing the motility of the two species. We note that this resembles crowd mosh pits~\cite{PhysRevLett.110.228701}, where an energetic minority drags less mobile neighbors at the boundary of an active region, though here the active region is not externally imposed but self-generated, nucleated by the persistent dopants themselves.

\begin{figure}[t]
\includegraphics[width=1\columnwidth]{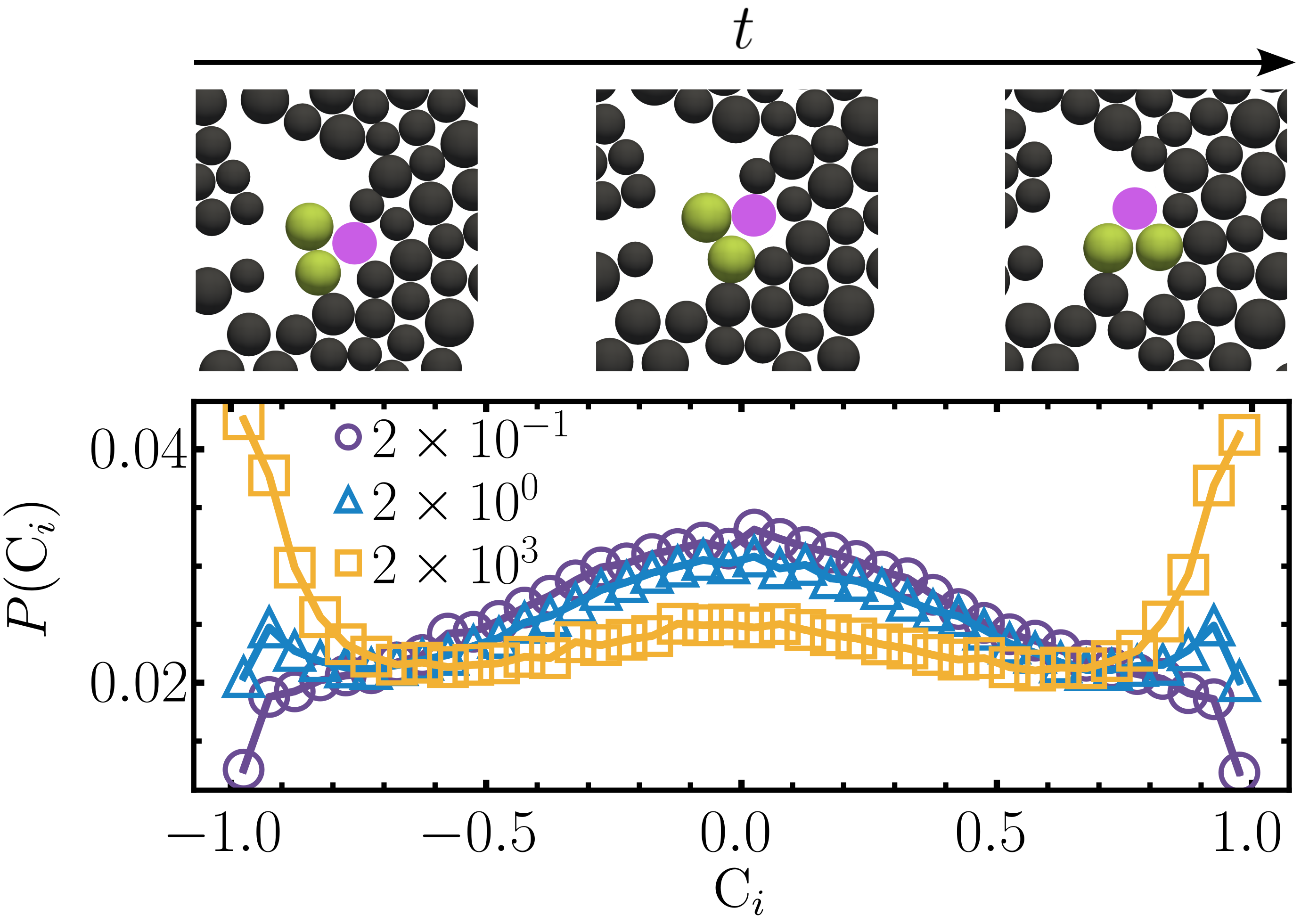}
\caption{Localized emergent chirality. Probability distribution of passive-particle chirality $P(\mathrm{C}_i)$ for three persistence lengths $l_p = 2\times 10^{-1}$ (purple circles), $2\times 10^{0}$ (blue triangles), and $2\times 10^{3}$ (yellow squares). At small persistence, $P(\mathrm{C}_i)$ is centered near zero, indicating no systematic local rotation. At large persistence, $P(\mathrm{C}_i)$ develops two symmetric peaks near $\mathrm{C}_i \simeq \pm 1$, revealing intermittent clockwise and counterclockwise rotations on timescales shorter than $\tau_r$. Time-ordered snapshots (separated by $2\Delta t$) illustrate a representative event: a passive particle (pink) undergoes a transient rotation while dragged by neighboring active particles (yellow); all other particles are shown in black.}
\label{fig:4} 
\end{figure}
As a dynamical by-product of boundary-localized transport, passive particles exhibit intermittent particle-scale rotations. In Fig.~\ref{fig:4}, we show that the homogeneous fluid has a chirality distribution centered at zero, whereas the void-forming regime develops symmetric peaks at finite $\mathrm{C}_i$, indicating transient clockwise and counterclockwise rotations. These events arise from persistent forcing under self-generated confinement, as persistent, misaligned propulsion along the void boundary produces local rotational displacements without global symmetry breaking. Importantly, this chiral motion is not an independent instability but a direct consequence of void formation in the presence of persistent propulsion. Consistent with this mechanism, local chirality in the void-forming regime is only apparent when quantified over time windows shorter than the persistence time $\tau_r$. While $l_p$ provides a convenient control parameter~\cite{PhysRevLett.127.278002} to distinguish the liquid-like regimes, identifying chiral events requires temporal resolution set by $\tau_r$ (see Supplemental Material~\cite{SI}). 

By contrast, a passive hot-cold mixture injects stress through Gaussian white forcing, which lacks temporal persistence and therefore cannot accumulate stress coherently; accordingly, we observe neither void formation nor chiral events in that case (see Supplemental Material~\cite{SI}).
Taken together, these results show that strong persistence does not merely amplify rearrangements; instead, it reorganizes the dynamics into a void-forming state with boundary-localized transport. In this regime, persistent propulsion breaks time-reversal symmetry at the particle level. Particles entering the void move ballistically over a persistence length $l_p$, and confinement geometrically rectifies this directional flux into collective rotation.

We have shown that active dopants do not simply melt a dense arrested medium. Sufficient persistence fundamentally alters the fluidization mechanism beyond an effective-temperature mapping. Stress injection exceeds relaxation, and dopants collectively amplify mechanically perturbed regions until their overlap triggers a localized instability. This nucleates self-generated voids and reorganizes transport into a boundary-confined mode. Within it, directed propulsion is converted into intermittent particle-scale rotations without global chiral order.
The underlying mechanism is captured by stress-diffusion and elastoplastic descriptions of driven amorphous solids~\cite{Nicolas2018,MatozPRL2017} and parallels recent evidence that spatially correlated forcing renormalizes mechanical response in disordered systems~\cite{morse2021direct,Agoritsas2024memory}. 

The distinctive feature here is that persistence dynamically generates the mechanically correlated region whose overlap drives the instability, rather than requiring externally imposed structure or chirality. This reveals a generic, persistence-controlled localization mechanism in disordered solids, with no direct analogue in the thermal limit.
We expect this mechanism to be relevant across a broad class of dense active-passive systems~\cite{RevModPhys.88.045006}, including driven colloids, granular media, and biological suspensions where activity is heterogeneous and structural relaxation is slow~\cite{grosser21,coban21}. More broadly, persistence-controlled stress accumulation may provide a design principle for disordered active materials with programmable transport localization or spatially selective mechanical response~\cite{grober2023unconventional}.

\emph{Acknowledgments.} DMF and GJ acknowledge support from MINECO (Grant No. PID2023-148991NA-I00). DMF is also funded through the Ramon y Cajal Program of the Spanish Ministry of Science and Innovation (Grant No. RYC2023-043002-I). GJ gratefully acknowledges the hospitality of the University of Lisbon during the early stages of this work. NAMA acknowledges support from the Portuguese Foundation for Science and Technology (FCT) under Contracts No.~UIDB/00618/2020 
(\href{https://doi.org/10.54499/UIDB/00618/2020}{10.54499/UIDB/00618/2020}), 
UIDP/00618/2020 
(\href{https://doi.org/10.54499/UIDP/00618/2020}{10.54499/UIDP/00618/2020}), 
and UID/00618/2025.

\emph{Data availability.} The data that support the findings of
this article are openly available~\cite{active_dopants_dataset_zenodo}. 
\bibliographystyle{apsrev4-1} 
\bibliography{./biblio}

\begin{thebibliography}{52}
\bibitem [{act()}]{active_dopants_dataset_zenodo}%
  \BibitemOpen
  \href@noop {} {}\bibinfo {note} {The dataset is available at [URL] and
  includes one trajectory for each data point in Fig.~2, along with all
  processed data used to reproduce the figures. The complete set of raw
  simulation outputs is available upon reasonable request.}\BibitemShut {Stop}%
\end{thebibliography}


\clearpage
\onecolumngrid

\setcounter{equation}{0}
\setcounter{figure}{0}
\setcounter{table}{0}
\renewcommand{\theequation}{S\arabic{equation}}
\renewcommand{\thefigure}{S\arabic{figure}}
\renewcommand{\thetable}{S\arabic{table}}

\section*{Supplemental Material}

\section{Minimal mechanical estimate for the void-forming onset: stress diffusion and dopant overlap criterion}
\subsection*{From Geometric Counting to Mechanical Influence: Scaling of Void Nucleation}

Void nucleation is a collective event: a single dopant produces a marginally subcritical stress dose, and irreversible rearrangement requires reinforcement by multiple statistically independent dopants within one persistence interval. To quantify this, we define the mean mechanical overlap number
\begin{equation}
\mathcal{N}_{\rm mech}(l_p)
\equiv
\rho_a \, V_{\rm infl}(l_p),
\end{equation}
which measures the expected number of independent dopants whose stress-influence regions overlap a typical spatial point during one persistence time.

Here, subcriticality refers to the typical stress dose generated by an isolated dopant in the dilute limit. Direct numerical inspection confirms that single dopants do not nucleate stable voids when $\rho_a \to 0$. While rare disorder configurations may locally exceed the yield threshold, such events do not produce sustained void growth and remain statistically negligible at low density. The onset considered here therefore marks the density at which reinforcement becomes typical rather than fluctuation-driven.

Thus, for $\mathcal{N}_{\rm mech} \ll 1$, stress events are isolated and the cumulative dose remains subcritical almost everywhere. In contrast, when
\begin{equation}
\rho_a \, V_{\rm infl}(l_p^\star) \sim O(1),
\label{eq:overlap_criterion}
\end{equation}
multi-dopant overlap becomes typical: fluctuations in the cumulative stress are no longer rare, and reinforcement above the yield threshold becomes statistically likely. 

Equation~\eqref{eq:overlap_criterion} should be interpreted as a mechanical overlap criterion rather than a statement of critical geometric percolation. It expresses that stress-influence regions become sufficiently dense that reinforcement is typical at a representative point in space. Unlike standard percolation transitions, no universal critical exponents are invoked; the instability reflects the competition between stress injection and relaxation.

We therefore focus on the regime where single dopants are subcritical and void formation emerges from the collective overlap of mechanically perturbed regions. This competition between local stress injection and structural relaxation is reminiscent of defect nucleation induced by a driven particle in crystalline or glassy media~\cite{Reichhardt2004PhysRevLettSI}, although here the forcing arises from persistent active dopants rather than external drive.

Before inserting the detailed mechanical response of the host, it is useful to establish a geometric upper bound. In a dilute, non-interacting medium, a persistent dopant explores a disk of radius $l_p = v_0 \tau_r$. The maximal region it can influence is therefore bounded by its ballistic exploration area,
\begin{equation}
    V_{\rm geom}(l_p) \sim l_p^2,
\end{equation}
which provides an upper scaling estimate: no mechanically mediated influence can exceed the region physically explored during a persistence run.

A purely ballistic overlap argument based on $V_{\rm geom}$ would predict a threshold scaling $l_p^\star \sim \phi_a^{-1/2}$, corresponding to geometric overlap of persistent trajectories. The observed inverse scaling instead indicates that the relevant influence volume is set by mechanically mediated stress propagation rather than direct trajectory overlap.

In a dense amorphous host, steric constraints limit ballistic exploration, while stress can propagate beyond direct contact through the surrounding matrix. The mechanically determined influence radius $r_{\rm infl}(l_p)$ therefore defines the effective interaction volume, which need not coincide with the purely geometric exploration area. In constrained media, one generically expects $V_{\rm infl}(l_p) \lesssim V_{\rm geom}(l_p)$, up to non-universal prefactors set by the host response.

\subsection*{Stress Diffusion and the Mechanical Influence Volume}

In this section, we derive the stress diffusion from overdamped force balance and a minimal viscoelastic constitutive relation. The goal is to justify the diffusion-relaxation equation used to define the mechanical influence region of a persistent dopant. At the coarse-grained level, inertia is negligible, and momentum balance reduces to
\begin{equation}
\nabla \cdot \boldsymbol{\sigma} - \zeta \mathbf{v}
+ \mathbf{f}_{\rm act} = 0 ,
\label{eq:force_balance}
\end{equation}
where $\boldsymbol{\sigma}$ is the stress tensor,
$\mathbf{v}$ the velocity field,
$\zeta$ an effective friction coefficient
(e.g., substrate or solvent drag),
and $\mathbf{f}_{\rm act}$ an active body-force density. Solving Eq.~(\ref{eq:force_balance}) for the velocity gives
\begin{equation}
\mathbf{v}
=
\frac{1}{\zeta}
\left(
\nabla \cdot \boldsymbol{\sigma}
+
\mathbf{f}_{\rm act}
\right).
\label{eq:velocity_eliminated}
\end{equation}

\subsubsection*{Maxwell viscoelastic constitutive relation.}

The derivation requires only diffusive stress propagation with
finite memory in an overdamped host. Under these conditions,
the accumulated stress during a persistence interval is controlled
by diffusive spreading and relaxation. The Maxwell model is used
as the minimal linear realization of these properties~\cite{Nicolas2018SI, Matoz2020SI},
\begin{equation}
\partial_t \boldsymbol{\sigma}
=
G\,\mathbf{E}
-
\frac{1}{\tau_m}\boldsymbol{\sigma},
\label{eq:maxwell}
\end{equation}
where $G$ is an effective shear modulus,
$\tau_m$ a stress relaxation time,
and
\(
\mathbf{E}
=
\frac{1}{2}
\left(
\nabla \mathbf{v}
+
(\nabla \mathbf{v})^{\!\top}
\right)
\)
is the strain-rate tensor. 

\subsubsection*{Elimination of the velocity field.}
Substituting Eq.~(\ref{eq:velocity_eliminated}) into the strain-rate definition yields, schematically,
\begin{equation}
\mathbf{E}
\sim
\nabla \mathbf{v}
=
\frac{1}{\zeta}
\left(
\nabla\nabla \cdot \boldsymbol{\sigma}
+
\nabla \mathbf{f}_{\rm act}
\right).
\label{eq:strain_rate}
\end{equation}
In an isotropic medium and at the level of scaling,
the operator $\nabla\nabla\cdot\boldsymbol{\sigma}$
renormalizes to a Laplacian acting on the stress tensor,
\begin{equation}
\nabla\nabla \cdot \boldsymbol{\sigma}
\;\rightarrow\;
\nabla^2 \boldsymbol{\sigma}.
\label{eq:isotropic_closure}
\end{equation}
The replacement of the anisotropic elastic propagator by a Laplacian
constitutes a coarse-grained closure valid at scales larger than the
typical rearrangement spacing. We emphasize that this approximation concerns the scaling of the radial envelope of the stress field and does not resolve the Eshelby quadrupolar structure of individual plastic rearrangements. In amorphous solids, plastic rearrangements redistribute stress
through long-range anisotropic Eshelby kernels
\cite{Nicolas2018SI}. The present diffusive closure retains
only the coarse-grained radial envelope of this redistribution.


Inserting Eq.~(\ref{eq:strain_rate}) into Eq.~(\ref{eq:maxwell}) then gives
\begin{equation}
\partial_t \boldsymbol{\sigma}
=
\frac{G}{\zeta}
\nabla^2 \boldsymbol{\sigma}
-
\frac{1}{\tau_m}\boldsymbol{\sigma}
+
\frac{G}{\zeta}
\nabla \mathbf{f}_{\rm act}.
\label{eq:stress_tensor_diffusion}
\end{equation}

\subsubsection*{Identification of coefficients.}

Define the stress diffusivity
\begin{equation}
D_\sigma \equiv \frac{G}{\zeta},
\label{eq:stress_diffusivity}
\end{equation}
which has dimensions of $L^2/T$.
Equation~(\ref{eq:stress_tensor_diffusion}) becomes
\begin{equation}
\partial_t \boldsymbol{\sigma}
=
D_\sigma \nabla^2 \boldsymbol{\sigma}
-
\frac{1}{\tau_m}\boldsymbol{\sigma}
+
D_\sigma \nabla \mathbf{f}_{\rm act}.
\label{eq:stress_tensor_final}
\end{equation}
For the purposes of estimating the spatial envelope of stress
magnitude, we coarse-grain the tensorial structure into a scalar
stress amplitude $\sigma(\mathbf r,t)$ and absorb the gradient of
the localized active force into a scalar source term
$s(\mathbf r,t)$. This yields the diffusion-relaxation equation
\begin{equation}
\partial_t \sigma
=
D_\sigma \nabla^2 \sigma
-
\frac{1}{\tau_m}\sigma
+
s(\mathbf r,t),
\label{eq:stress_diffusion}
\end{equation}
which represents the spatially resolved counterpart of
mean-field stress-diffusion descriptions developed in
elastoplastic models of driven amorphous solids
\cite{Nicolas2018SI,MatozPRL2017SI}.

\subsection*{Connection to Stress Diffusion and the Collective Onset Criterion}

The minimal continuum description of the overdamped host Eq.~\eqref{eq:stress_diffusion} treats the local stress magnitude $\sigma(\mathbf r,t)$ as evolving according to a diffusion-relaxation equation with a finite-duration source representing the persistent dopant,

\begin{equation}
\partial_t \sigma
=
D_\sigma \nabla^2 \sigma
-
\frac{1}{\tau_m}\sigma
+
s(\mathbf r,t),
\label{eq:stress_diffusion}
\end{equation}
where $D_\sigma$ is a stress diffusivity and $\tau_m$ a stress relaxation time. For a localized source active over a time $\tau_r$, the region where the time-integrated stress exceeds a yield threshold defines the influence area.

We derive the scaling of the influence area $V_{\rm infl}$ for a single persistent dopant by solving the diffusion-relaxation equation~(\ref{eq:stress_diffusion}) with a finite-duration localized source and evaluating the time-integrated stress (``stress dose'').

\subsubsection*{Model and source.}
We consider the scalar stress magnitude $\sigma(\mathbf r,t)$ evolving in two dimensions according to

\begin{equation}
\partial_t \sigma(\mathbf r,t)
=
D_\sigma \nabla^2 \sigma(\mathbf r,t)
-
\frac{1}{\tau_m}\sigma(\mathbf r,t)
+
s(\mathbf r,t),
\label{eq:stress_pde_appendix}
\end{equation}
with initial condition $\sigma(\mathbf r,0)=0$. A single dopant located at the origin injects stress at a constant rate $\sigma_0$ for a time $\tau_r$,

\begin{equation}
s(\mathbf r,t)
=
\sigma_0\,\delta(\mathbf r)\,H(t)\,H(\tau_r-t),
\label{eq:source_appendix}
\end{equation}
where $H(t)$ is the Heaviside step function. The product $H(t)H(\tau_r-t)$ restricts the source to a finite time window of duration $\tau_r$, representing a single persistence interval of the active dopant,
\begin{equation*}
    H(t)H(\tau_r - t)
=
\begin{cases}
1, & 0 < t < \tau_r, \\
0, & \text{otherwise}.
\end{cases}
\end{equation*}

\subsubsection*{Green's function.}
Let $G(\mathbf r,t)$ denote the Green's function of the homogeneous operator in Eq.~(\ref{eq:stress_pde_appendix}), i.e.,

\begin{equation}
\partial_t G(\mathbf r,t)
=
D_\sigma \nabla^2 G(\mathbf r,t)
-
\frac{1}{\tau_m}G(\mathbf r,t),
\qquad
G(\mathbf r,0^+)=\delta(\mathbf r).
\label{eq:greens_def}
\end{equation}

In $d=2$, translational invariance and standard Fourier methods yield

\begin{equation}
G(r,t)
=
\frac{1}{4\pi D_\sigma t}
\exp\!\left(-\frac{r^2}{4D_\sigma t}\right)
\exp\!\left(-\frac{t}{\tau_m}\right),
\qquad t>0,
\label{eq:greens_2d}
\end{equation}

where $r=|\mathbf r|$.

\subsubsection*{Solution by convolution.}
The solution of Eq.~(\ref{eq:stress_pde_appendix}) with $\sigma(\mathbf r,0)=0$ is the space-time convolution

\begin{equation}
\sigma(\mathbf r,t)
=
\int_0^t dt' \int d^2 r'\,
G(\mathbf r-\mathbf r',t-t')\,s(\mathbf r',t').
\label{eq:convolution}
\end{equation}

Substituting the point source~(\ref{eq:source_appendix}) gives

\begin{equation}
\sigma(r,t)
=
\sigma_0\int_0^{\min(t,\tau_r)} dt'\, G\!\left(r,t-t'\right),
\label{eq:sigma_min}
\end{equation}
where we have collapsed the spatial integral $s(r) \propto \delta(\mathbf r)$, so we simply evaluate the Green's function at $r$.

For times $0<t\le \tau_r$ (the only range needed below), Eq.~(\ref{eq:sigma_min}) reduces to

\begin{equation}
\sigma(r,t)
=
\sigma_0\int_0^{t} du\, G(r,u),
\qquad 0<t\le\tau_r,
\label{eq:sigma_tle}
\end{equation}
where we changed variables to $u=t-t'$.

\subsubsection*{Stress dose and exchange of integrations.}
Define the stress dose accumulated during one persistence interval:

\begin{equation}
\mathcal I(r)\equiv \int_0^{\tau_r} dt\, \sigma(r,t).
\label{eq:dose_def}
\end{equation}
Substituting Eq.~(\ref{eq:sigma_tle}) and exchanging the order of integration yields

\begin{align}
\mathcal I(r)
&=
\sigma_0 \int_0^{\tau_r} dt \int_0^{t} du\, G(r,u)
\nonumber\\
&=
\sigma_0 \int_0^{\tau_r} du\, (\tau_r-u)\,G(r,u).
\label{eq:dose_exchange}
\end{align}
Using Eq.~(\ref{eq:greens_2d}) gives the explicit representation

\begin{equation}
\mathcal I(r)
=
\frac{\sigma_0}{4\pi D_\sigma}
\int_0^{\tau_r} du\,
\frac{\tau_r-u}{u}\,
\exp\!\left(-\frac{r^2}{4D_\sigma u}\right)\exp\!\left(-\frac{u}{\tau_m}\right).
\label{eq:dose_integral_exact}
\end{equation}

\subsubsection*{Role of the stress relaxation time $\tau_m$.}

The stress field undergoes both spatial diffusion and local relaxation.
The relaxation time $\tau_m$ controls how long stress injected by an
isolated dopant persists before being dissipated by plastic or
viscoelastic processes. For a single dopant, two regimes can be
distinguished depending on the ratio $\tau_r/\tau_m$:

\begin{description}
    \item[(i)] \emph{Long persistence:} $\tau_r \gg \tau_m$. 
    When the persistence time exceeds the intrinsic relaxation time,
    stress injected at early times decays before the run ends.
    The exponential damping factor $\exp(-t/\tau_m)$ effectively
    cuts off the time integral at $t \sim \tau_m$, leading to a
    saturation of the influence length at
    \begin{equation}
    r_{\rm infl} \sim \sqrt{D_\sigma \tau_m}.
    \end{equation}
    In this regime, the influence volume of an isolated dopant
    is limited by the host relaxation properties rather than
    by persistence.

    \item[(ii)] \emph{Short or comparable persistence:} $\tau_r \lesssim \tau_m$. 
    In this regime, stress does not significantly relax during
    a persistence interval. The exponential factor in the Green's
    function remains close to unity for $t \le \tau_r$, so that
    stress spreading is controlled primarily by diffusion,
    yielding $r_{\rm infl} \sim \sqrt{D_\sigma \tau_r}$.
\end{description}

Importantly, the discussion above applies to isolated dopants.
Near the onset of the void transition, overlapping dopants repeatedly
inject stress into the same spatial region. In this collective regime,
relaxation is dynamically compensated by renewed forcing, so that
stress accumulation can exceed the single-dopant relaxation limit.
The instability therefore emerges when collective reinforcement
overcomes relaxation.

\subsubsection*{Asymptotic regime $\tau_r \lesssim \tau_m$: diffusive stress}

For persistence times not exceeding the stress relaxation time, $\tau_r \lesssim \tau_m$, the exponential relaxation factor satisfies $\exp(-u/\tau_m)\approx 1$ over $u\in[0,\tau_r]$, and Eq.~(\ref{eq:dose_integral_exact}) simplifies to

\begin{equation}
\mathcal I(r)
\approx
\frac{\sigma_0}{4\pi D_\sigma}
\int_0^{\tau_r} du\,
\frac{\tau_r-u}{u}\,
\exp\!\left(-\frac{r^2}{4D_\sigma u}\right),
\label{eq:dose_integral_simplified}
\end{equation}
which represents the diffusive limit. 

\subsubsection*{Reduction to exponential-integral functions.}
Introduce the dimensionless variable

\begin{equation}
x \equiv \frac{r^2}{4D_\sigma u}
\quad \Longleftrightarrow \quad
u=\frac{r^2}{4D_\sigma x},
\qquad
du=-\frac{r^2}{4D_\sigma}\frac{dx}{x^2}.
\label{eq:change_variable}
\end{equation}
The variables exchange bounds $u\to 0^+$ to $x\to \infty$, and $u=\tau_r$ to

\begin{equation}
x_r \equiv \frac{r^2}{4D_\sigma \tau_r}.
\label{eq:xr_def}
\end{equation}
Using Eq.~(\ref{eq:change_variable}), Eq.~(\ref{eq:dose_integral_simplified}) becomes

\begin{align}
\mathcal I(r)
&\approx
\frac{\sigma_0}{4\pi D_\sigma}
\int_{x_r}^{\infty} dx\,
\left[
\tau_r\frac{e^{-x}}{x}
-
\frac{r^2}{4D_\sigma}\frac{e^{-x}}{x^2}
\right].
\label{eq:dose_x_form}
\end{align}
Define the exponential integral $E_1(z)$,

\begin{equation}
E_1(z)\equiv \int_z^\infty \frac{e^{-x}}{x}\,dx,
\label{eq:E1_def}
\end{equation}
and note the identity

\begin{equation}
\int_z^\infty \frac{e^{-x}}{x^2}\,dx
=
\frac{e^{-z}}{z}-E_1(z),
\label{eq:identity_x2}
\end{equation}
which follows from integration by parts. Equation~(\ref{eq:dose_x_form}) then yields

\begin{equation}
\mathcal I(r)
\approx
\frac{\sigma_0}{4\pi D_\sigma}
\left[
\tau_r E_1(x_r)
-
\frac{r^2}{4D_\sigma}\left(\frac{e^{-x_r}}{x_r}-E_1(x_r)\right)
\right],
\qquad x_r=\frac{r^2}{4D_\sigma\tau_r}.
\label{eq:dose_E1_exact}
\end{equation}

\subsubsection*{Near-field asymptotics and influence radius.}
We now analyze the regime $r \ll \sqrt{D_\sigma \tau_r}$(i.e.,\ $x_r\ll 1$).
This condition corresponds to distances smaller than the diffusive
length reached by stress during a persistence interval. The small-argument expansion reads,

\begin{equation}
E_1(x)= -\gamma_E - \ln x + x + O(x^2),
\label{eq:E1_small}
\end{equation}
where $\gamma_e \approx 0.57$ is the Euler-Mascheroni constant. Now, up to $O(1)$ constants,

\begin{equation}
\mathcal I(r)
\sim
\frac{\sigma_0 \tau_r}{4\pi D_\sigma}
\left[
\ln\!\left(\frac{4D_\sigma \tau_r}{r^2}\right) + O(1)
\right],
\qquad r\ll \sqrt{D_\sigma \tau_r}.
\label{eq:dose_log}
\end{equation}
Next, we define the influence radius $r_{\rm infl}$ as the largest distance from the active dopant at which the accumulated stress dose during one persistence interval is sufficient to trigger an irreversible local rearrangement. Formally, $r_{\rm infl}$ is determined by

\begin{equation}
\mathcal I(r_{\rm infl})=\Sigma_y.
\label{eq:rinfl_def}
\end{equation}
The yield dose $\Sigma_y$ is defined as the minimal time-integrated
stress required to trigger an irreversible local rearrangement.
It serves as a coarse-grained threshold for single-dopant cage
destabilization. We focus on the regime in which the dose generated
by an isolated dopant remains subcritical, so that void formation
emerges from collective overlap. Such coarse-grained yielding
criteria are standard in elastoplastic descriptions of amorphous
solids \cite{Nicolas2018SI}; here $\Sigma_y$ extends this concept
to stress accumulated over a finite persistence interval.

Thus, using Eq.~\eqref{eq:dose_log} gives

\begin{equation}
\Sigma_y
\sim
\frac{\sigma_0 \tau_r}{4\pi D_\sigma}
\left[
\ln\!\left(\frac{4D_\sigma \tau_r}{r_{\rm infl}^2}\right) + O(1)
\right],
\label{eq:rinfl_equation}
\end{equation}

\subsubsection*{From yielding stress to a radius of influence}

First we write the $O(1)$ term as an unknown dimensionless constant $c_0$,
\begin{equation*}
\Sigma_y =
\frac{\sigma_0 \tau_r}{4\pi D_\sigma}
\left[
\ln\!\left(\frac{4D_\sigma \tau_r}{r_{\rm infl}^2}\right) + c_0
\right].
\end{equation*}
Here, $c_0$ absorbs cutoff-dependent and nonuniversal contributions. Now rearranging, 

\begin{equation}
\ln\!\left(\frac{4D_\sigma \tau_r}{r_{\rm infl}^2}\right)
=
\frac{4\pi D_\sigma}{\sigma_0}\frac{\Sigma_y}{\tau_r}
- c_0,
\end{equation}
which implies

\begin{equation}
r_{\rm infl}^2
\sim
4D_\sigma \tau_r
\,
\exp\!\left[-\frac{4\pi D_\sigma}{\sigma_0}\frac{\Sigma_y}{\tau_r}
+ O(1)\right].
\label{eq:rinfl_solution}
\end{equation}
Equation~\eqref{eq:rinfl_solution} shows that the influence radius
is controlled primarily by the diffusive length
$\sqrt{D_\sigma \tau_r}$, with an exponential correction that encodes the finite yielding threshold. For sufficiently persistent
forcing, such that
\(
\frac{4\pi D_\sigma}{\sigma_0}\frac{\Sigma_y}{\tau_r}
= O(1)
\)
or smaller, the exponential factor remains of order unity, and the dominant scaling reduces to
\begin{equation*}
r_{\rm infl}^2 \sim D_\sigma \tau_r,
\label{eq:rinfl_scaling_derived}
\end{equation*}
up to logarithmic corrections characteristic of two-dimensional diffusion. The area of influence is then,
\begin{equation}
    V_{\rm infl}\sim \pi r_{\rm infl}^2 \sim D_\sigma \tau_r,
    \label{eq:vinfl_scaling_derived}
\end{equation}
Now, since $l_p = v_0 \tau_r$ at fixed propulsion speed $v_0$, Eq.~(\ref{eq:vinfl_scaling_derived}) implies
\begin{equation}
V_{\rm infl} \propto l_p.
\end{equation}
\begin{figure}[h]
    \centering
    \includegraphics[width=0.85\linewidth]{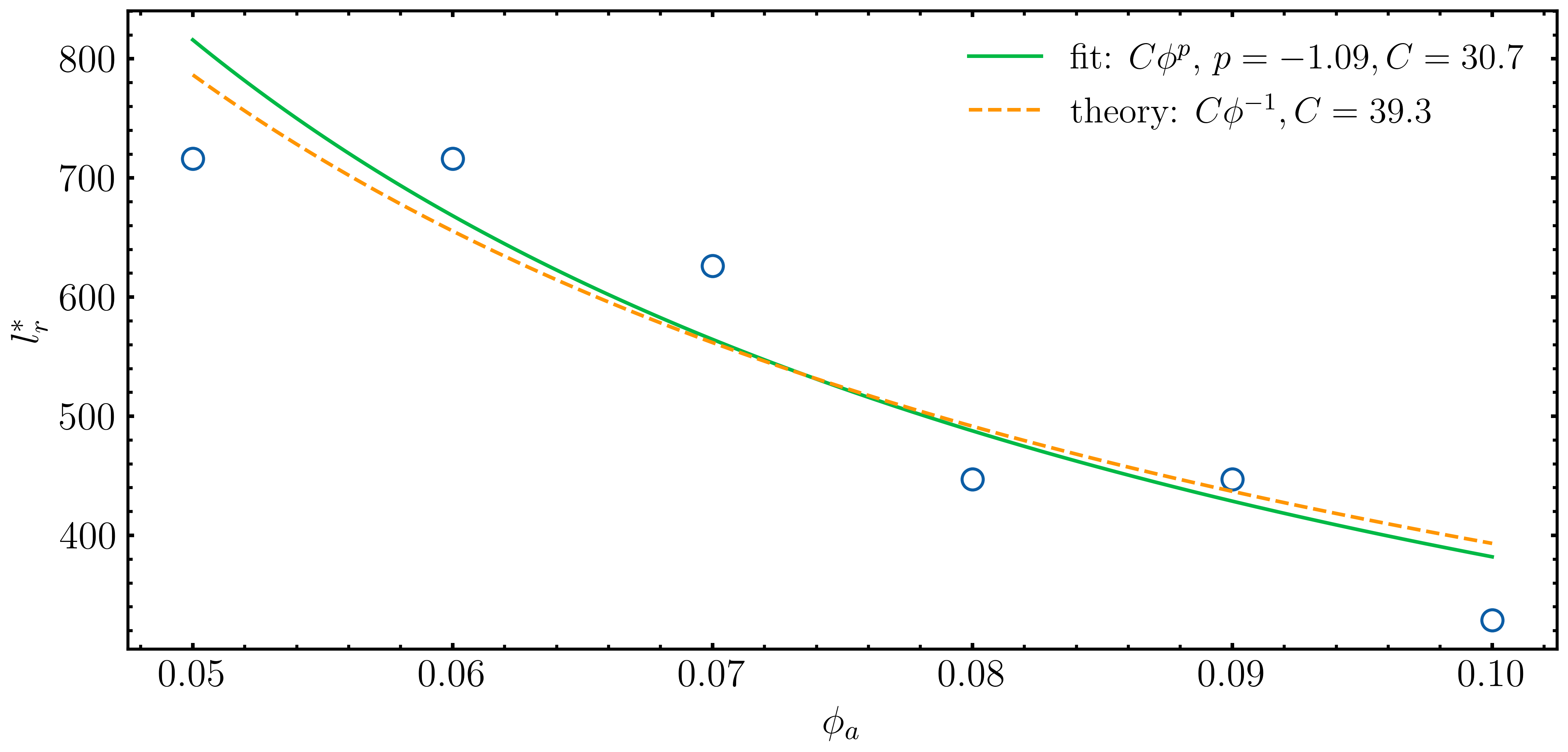}
    \caption{Void-onset persistence length $l_p^\star$ as a function of the active fraction $\phi_a$, determined from the phase boundaries shown in Fig.~2 of the main text. Symbols with error bars denote numerical measurements; error bars reflect statistical uncertainty in determining the onset boundary. The solid line shows the continuum prediction $l_p^\star \propto \phi_a^{-1}$, while the dashed line corresponds to a linear fit to the data. The agreement supports the stress-diffusion scaling $V_{\rm infl} \sim D_\sigma \tau_r$ and the resulting collective overlap criterion.}
    \label{fig:void_onset}
\end{figure}
This scaling replaces the geometric estimate $V_{\rm infl} \sim l_p^2$ and reflects the fact that mechanical influence is controlled by diffusive stress propagation rather than ballistic exploration. Finally, substituting into the collective onset condition
$\rho_a V_{\rm infl}(l_p^\star) \sim 1$,
and using $\rho_a \propto \phi_a$ at fixed packing fraction, yields
\begin{equation}
l_p^\star \sim \phi_a^{-1}.
\label{eq:lp_phi_final}
\end{equation}
We emphasize that this scaling reflects a mechanical instability criterion rather than purely geometric percolation. In overdamped systems, each persistent dopant injects stress over a duration $\tau_r$, generating a mechanically perturbed region of size $r_{\rm infl} \sim \sqrt{D_\sigma \tau_r}$. When the typical dopant spacing $\rho_a^{-1/2}$ becomes comparable to $r_{\rm infl}$, stress fields overlap before dissipating, enabling collective reinforcement to exceed the local yield threshold. This contrasts with ballistic geometric overlap of persistent trajectories, which would predict $l_p^\star \sim \phi_a^{-1/2}$. The observed inverse scaling therefore identifies stress accumulation, rather than kinematic overlap, as the operative mechanism.

To test this prediction, we fit the numerically determined onset
boundary to Eq.~\eqref{eq:lp_phi_final}
(see Fig.~\ref{fig:void_onset}). The data are consistent with an
inverse dependence of $l_p^\star$ on $\phi_a$ within numerical
uncertainty. A best-fit slope agrees with the continuum prediction
up to a non-universal prefactor, as expected from the coarse-grained
treatment of stress diffusion and the neglect of microscopic host
heterogeneity. Importantly, the inferred onset boundary does not depend on how it is measured. We obtain the same boundary both from fluctuations of the mean rearrangement fraction and from direct measurements of the void area (Sec.~\ref{sec:void_char}).

This agreement identifies void nucleation as a mechanically mediated collective instability governed by diffusive stress propagation, and extends mean-field stress-diffusion approaches to a spatially resolved active regime~\cite{MatozPRL2017SI}.
Persistent dopants act as spatially correlated local
forcing, conceptually related to the AQRD protocol of
Ref.~\cite{Agoritsas2024memorySI}. There, cooperativity
renormalizes the effective modulus and controls plastic
activity; here, persistence sets the spatial range of
stress injection and thereby governs the onset of instability.
Unlike quasistatic mappings between shear and random forcing
\cite{morse2021directSI}, the present mechanism relies on
finite persistence and time-accumulated stress, identifying
a dynamical route to void formation.
\section{Additional measurements}
\begin{figure}[h]
    \centering
\includegraphics[width=0.45\linewidth]{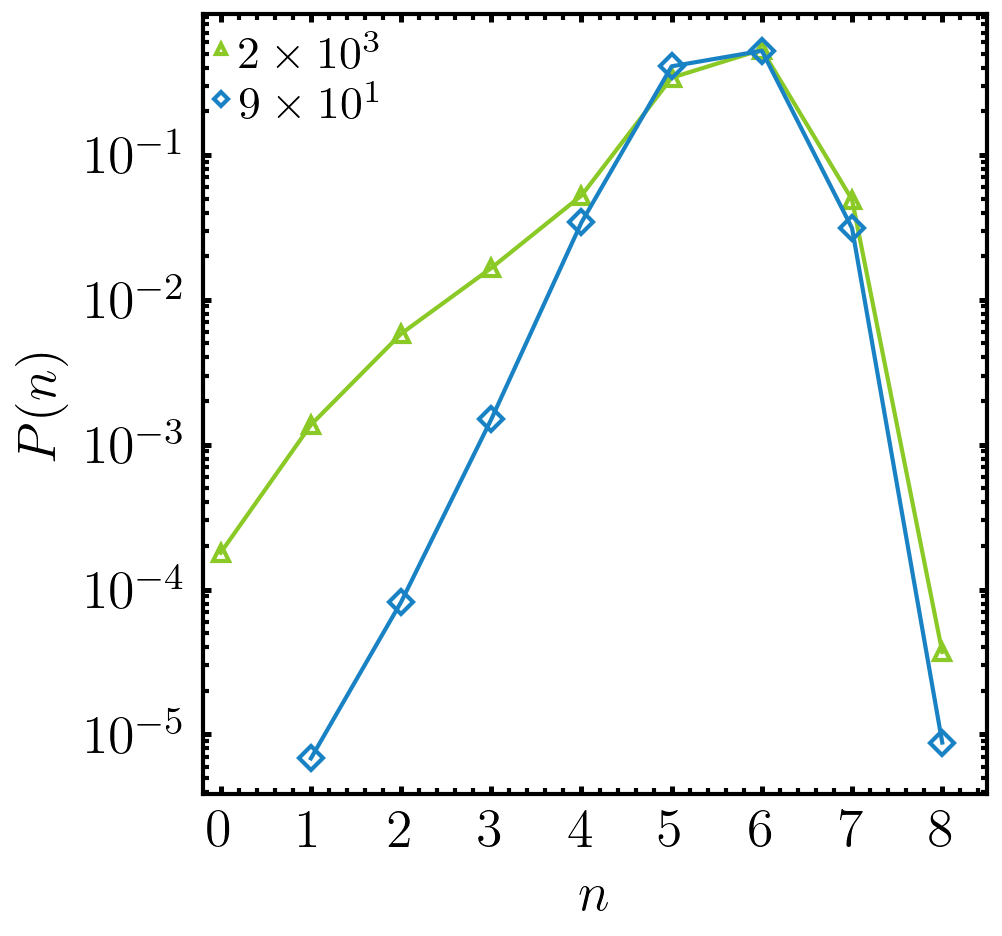}
    \caption{Probability distribution of the coordination number (number of nearest neighbors), $P(n)$, shown with a logarithmic $y$ axis, at $\phi_a=10^{-1}$ for two persistence lengths: $l_p=9\times10^{1}$ (green triangles) and $l_p=2\times10^{3}$ (blue diamonds). At larger $l_p$, $P(n)$ has enhanced weight at low coordination, indicating an increased probability of particles with fewer than four neighbors compared to the less-persistent case.}
    \label{fig:S1}
\end{figure}
Here we present additional measurements not shown in the main text. These results provide further characterization of the fluid regimes and corroborate the main conclusions.
\subsection{Local coordination number}
To understand the origin of the non-monotonic behavior of the rearrangement fluctuations reported in the main text, we analyze here the distribution of the local coordination number $n$ (nearest-neighbor count) at intermediate and large persistence. Nearest neighbors are identified using a cutoff distance equal to twice the radius of the largest particle in the system. Figure~\ref{fig:S1} shows that, in both cases, the distribution peaks near $n\simeq 6$, as expected for a dense two-dimensional packing. In the intermediate-persistence case (the homogeneous-fluid regime), the probability of finding particles with fewer than four neighbors is small. By contrast, at large $l_p$ the distribution develops enhanced weight at low coordination, i.e., an increased probability of particles with $n<4$.
\begin{figure}[h]
    \centering
\includegraphics[width=0.6\linewidth]{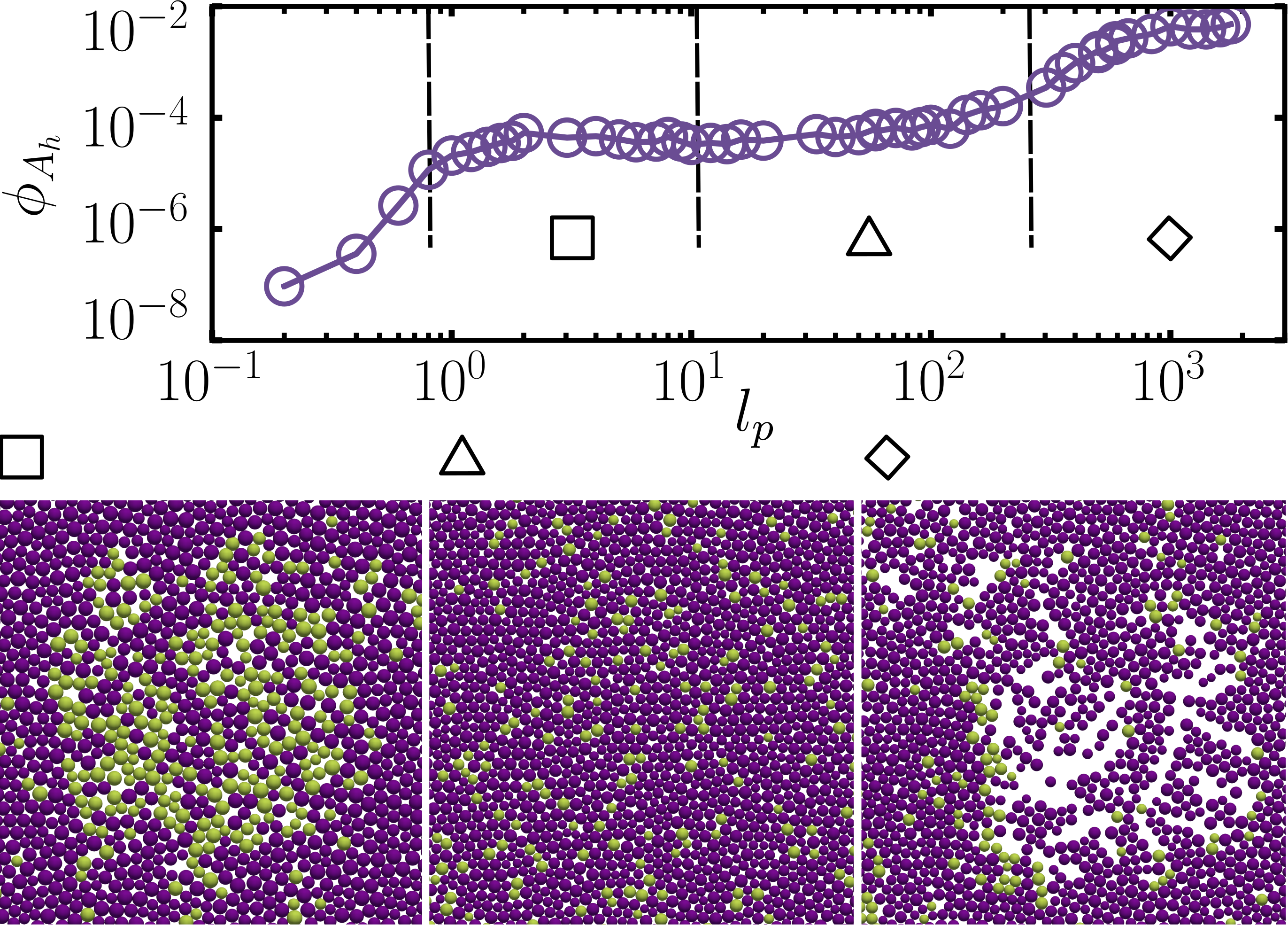}
    \caption{Largest-void area fraction $\phi_{A_h}$ as a function of the persistence length $l_p$ at $\phi_a=10^{-1}$. Snapshots show passive particles (purple) and active particles (yellow).}
    \label{fig:S2}
\end{figure}
\subsection{Explicit void characterization}
\label{sec:void_char}
To quantify voids, in each snapshot we draw a disk around every particle (with radius proportional to the particle size) and define the filled region as the union of all disks. Voids are then the connected empty regions inside the simulation box, i.e., the complement of the filled region. In the void-forming regime, the empty space can appear as many nearby holes separated by some particles, so we identify the largest connected void via a clustering procedure. Specifically, once the void region is obtained, we insert tracer points (implemented as very small particles) and compute the cluster-size distribution of these tracers. The area of the largest tracer cluster is then taken as the area of the largest void in the system.

Figure~\ref{fig:S2} shows the largest-void area fraction $\phi_{A_h}$, defined as the area of the largest void cluster divided by the simulation-box area, as a function of the persistence length at fixed $\phi_a=10^{-1}$. Consistent with the structural signatures in Fig.~1 of the main text, $\phi_{A_h}$ exhibits a non-monotonic dependence on persistence. It increases near the arrest-fluidization transition, remains approximately constant over an intermediate range of $l_p$, and then rises again upon entering the void-forming regime at large persistence.
Close to the onset of fluidization, we observe a state reminiscent of hot-cold passive mixtures~\cite{FreyPRL2016SI,ManningPRL2024SI}: the system demixes into an active region and a passive region. As shown in Fig.~\ref{fig:S2} (square symbol), the active particles form a clustered, mobile domain, while the passive particles remain largely caged in the surrounding medium. This is consistent with an effective-attraction scenario. In the regime where the persistence length of the active particles is small, the system reduces to Brownian particles of higher diffusivity ('hot') in a bath of lower-diffusivity particles ('cold'); in this limit, the active ('hot') particles effectively attract one another, promoting clustering~\cite{PhysRevFluids.2.043103SI}.

\begin{figure}[h]
\includegraphics[width=0.45\linewidth]{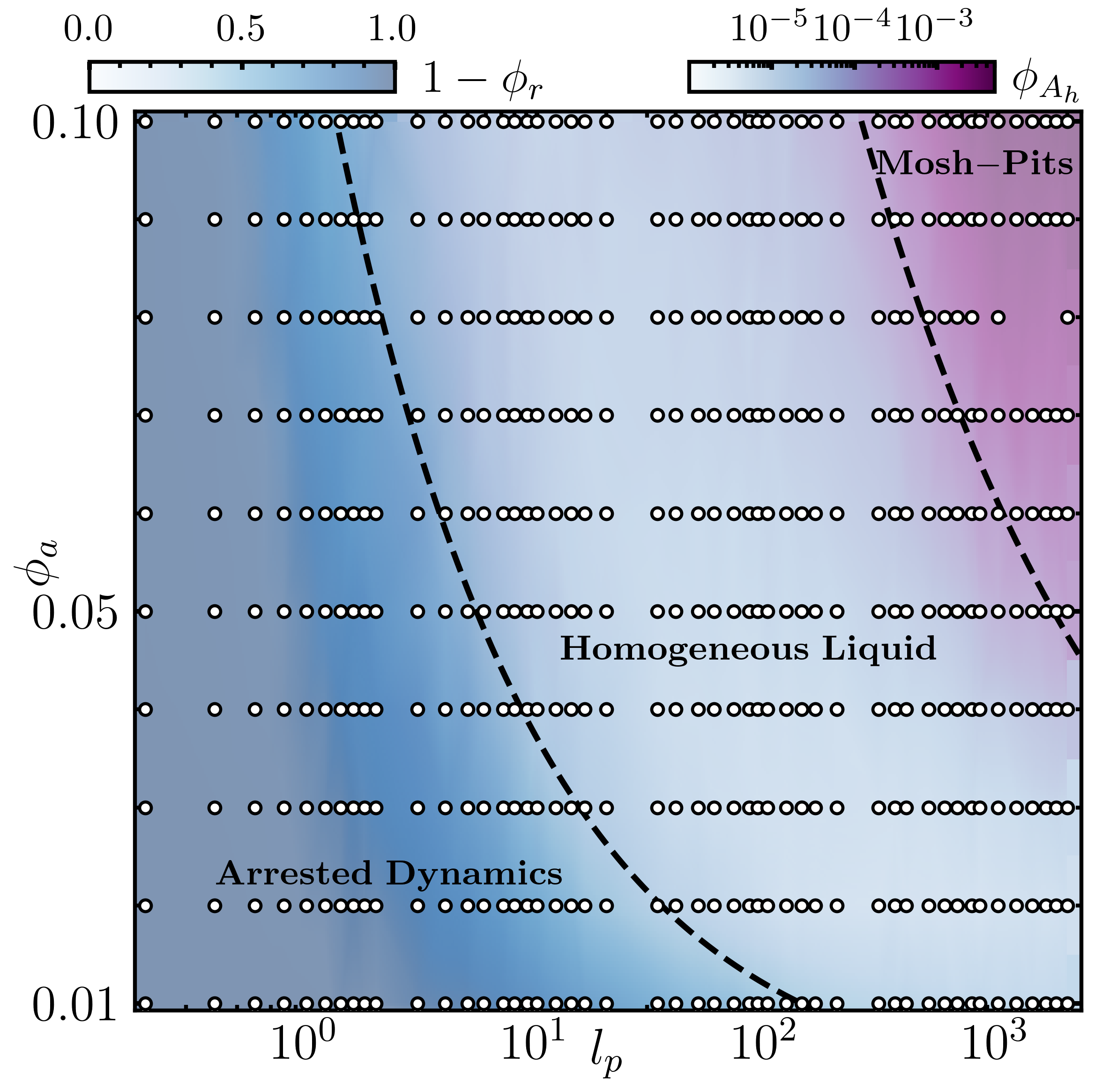}
\caption{Phase diagram of the active-passive mixture in the $(l_p,\phi_a)$ plane. The blue-white color scale shows $1-\phi_r$, where $\phi_r$ is the fraction of rearranging particles (dark blue: arrested; white: fluid). The light-blue-purple color scale shows the area fraction of the largest void, $\phi_{A_h}$, with deeper purple indicating larger voids. Black dashed lines mark the phase boundaries reported in Fig.~2 of the main text.}\label{fig:S3}
\end{figure}

Finally, to verify that the phase boundaries extracted from the non-monotonic behavior of $\chi_r$ reliably distinguish the arrested phase, the homogeneous fluid, and the void-forming regime, Fig.~\ref{fig:S3} shows the phase diagram (see Fig.~2 of the main text) with colors representing the rearrangement fraction $\phi_r$ and the area fraction of the largest void, $\phi_{A_h}$. The black dashed lines indicate the phase boundaries reported in Fig.~2 of the main text. The excellent agreement confirms that an explicit void characterization yields the same results as the criterion based on the rearrangement fluctuations $\chi_r$.

\subsection{Dependence on active dopant fraction}
\begin{figure}[h]
\includegraphics[width=0.55\columnwidth]{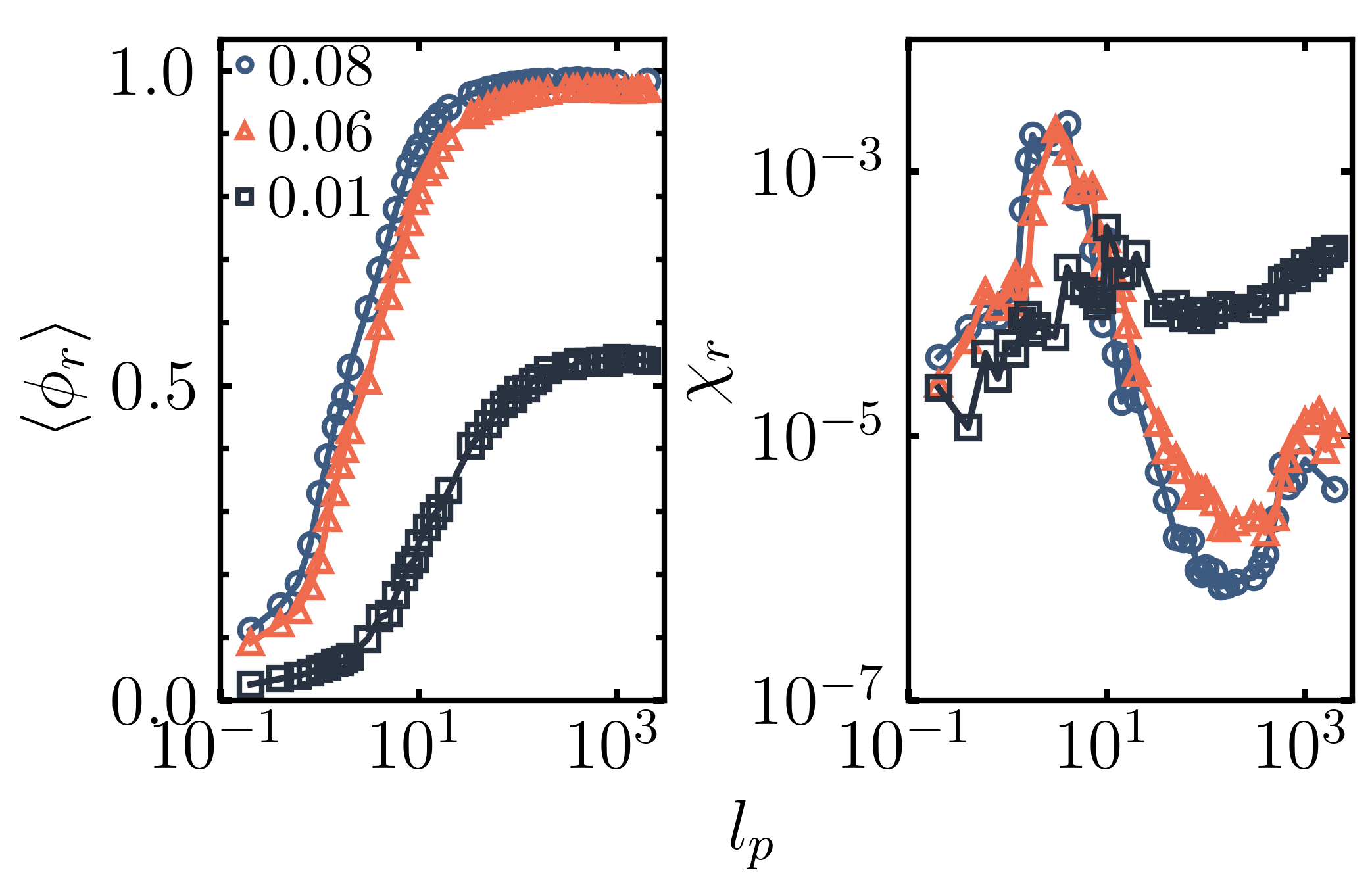}
\caption{Characterization of the system at fixed active fraction as a function of the persistence length $l_p$ for three active dopant fractions: $\phi_a=0.08$ (blue circles), $0.06$ (red triangles), and $0.01$ (black squares). (a) Mean rearrangement fraction, $\langle \phi_r \rangle$. (b) Rearrangement fluctuations, $\langle \phi_r^2 \rangle-\langle \phi_r \rangle^2$.} 
\label{fig:S4} 
\end{figure}

Figure~\ref{fig:S4} shows the mean number of rearrangements and their fluctuations as a function of the persistence length for different fractions of active dopants. For dopant fractions $\phi_a>0.05$, the mean number of rearrangements is approximately the same, and the fluctuations $\chi_r$ exhibit the same non-monotonic dependence on $l_p$. This similarity arises because, in this regime, the system enters the void-forming regime at large $l_p$.
By contrast, at smaller $\phi_a$, fewer rearrangements occur within the same observation window, and $\chi_r$ no longer displays a non-monotonic trend. This is consistent with the absence of a void-forming regime at very low dopant fractions.

\begin{figure}[h]
\includegraphics[width=0.45\columnwidth]{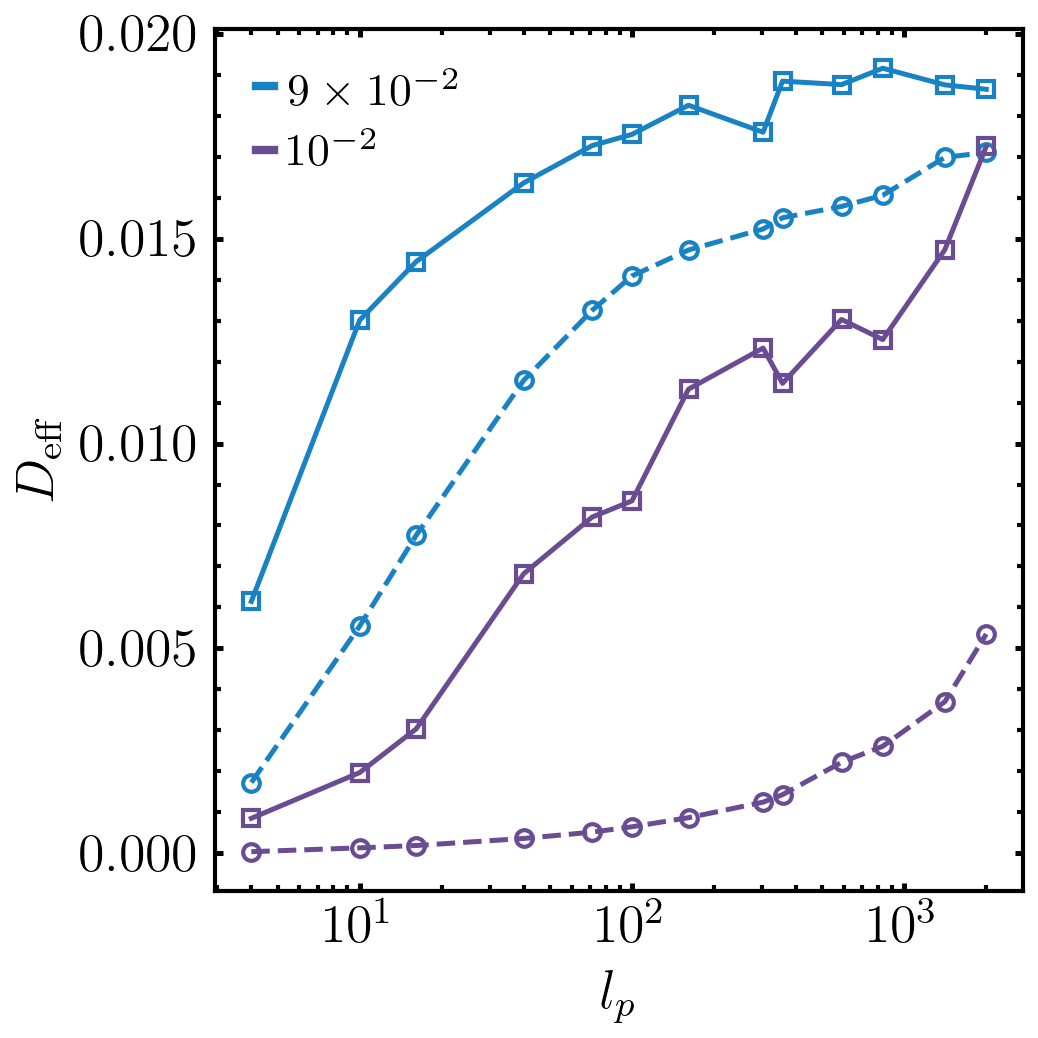}
\caption{Effective diffusion coefficient $D_{\mathrm{eff}}$ as a function of the persistence length $l_p$ for two active fractions, $\phi_a=9\times10^{-2}$ (blue) and $\phi_a=10^{-2}$ (purple). Squares and dashed lines denote the active-particle diffusivity, $D_{\mathrm{eff}}^{a}$, while circles and solid lines denote the passive-particle diffusivity, $D_{\mathrm{eff}}^{p}$.} 
\label{fig:S5} 
\end{figure}

Moreover, Fig.~\ref{fig:S5} shows the effective diffusion coefficients of passive and active particles for two dopant fractions. At low $\phi_a$, passive particles diffuse more slowly than active ones; however, for both species, the effective diffusion increases with persistence length $l_p$. At higher $\phi_a$, increasing $l_p$ again enhances the mobility of both species, and in the void-forming regime (very large $l_p$) the difference between the diffusion coefficients of active and passive particles is substantially reduced.

\subsection{Chirality}
\begin{figure}[t]
    \centering
\includegraphics[width=0.7\linewidth]{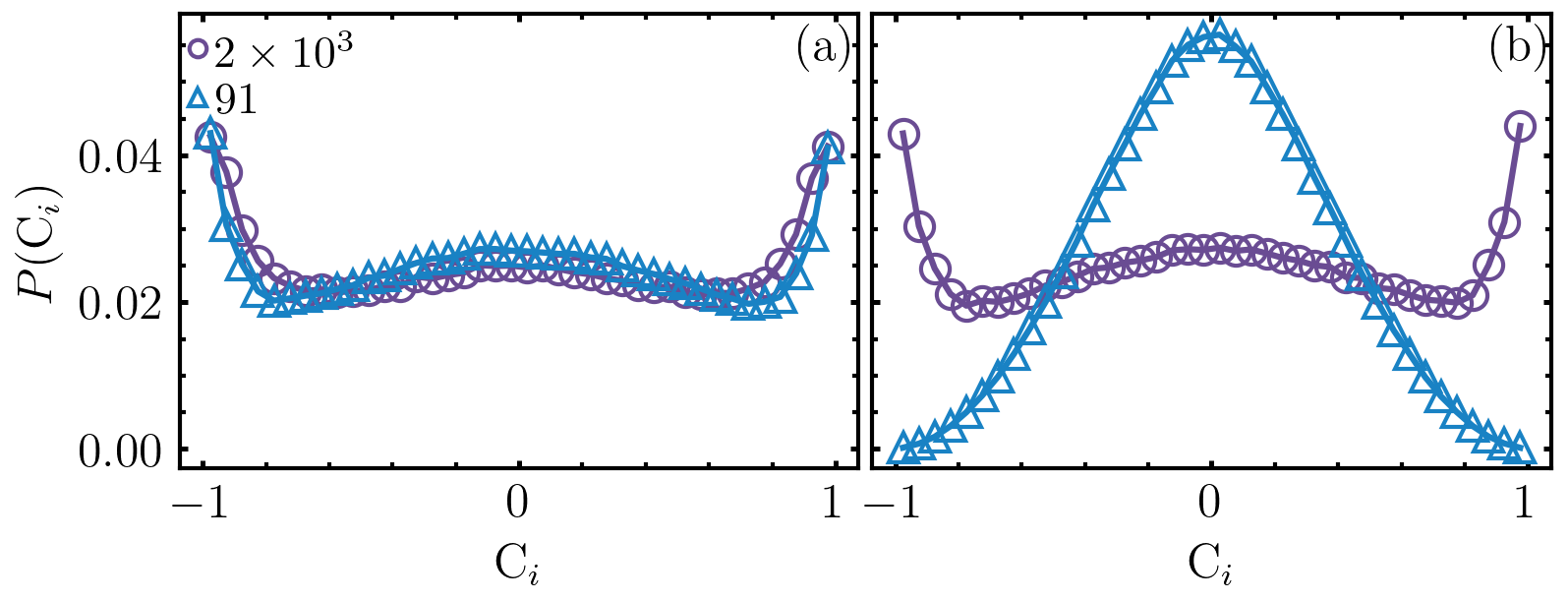}
    \caption{Probability distributions of the passive-particle chirality, $P(\mathrm{C}_i)$, at fixed persistence length $l_p = 2\times10^{3}$ for $dt = 5\times10^{-3}$ (left) and $dt = 10^{2}$ (right). Blue triangles correspond to dopants with $\tau_r=91$, while purple circles correspond to $v_0=1$ and $\tau_r=2\times10^{3}$. Local chirality is resolved only for $dt<\tau_r$.}
    \label{fig:S6}
\end{figure}

Figure~\ref{fig:S6} compares two systems with the same persistence length $l_p$ but different values of $v_0$ and $\tau_r$. The figure shows that, although $l_p$ is a convenient control parameter for recovering the different liquid-like regimes, identifying local chirality requires resolving the persistence time $\tau_r$, which sets the relevant temporal resolution. Specifically, when the time window used to compute chirality is shorter than the persistence time, both systems exhibit chiral signatures in the passive particles (Fig.~\ref{fig:S6}(a)). In contrast, when the time window exceeds $\tau_r$, chiral events are no longer detected, as indicated by a symmetric distribution centered at zero (Fig.~\ref{fig:S6}(b)).
\section{Hot-Cold Mixtures}
\begin{figure}[h]
    \centering
\includegraphics[width=0.45\linewidth]{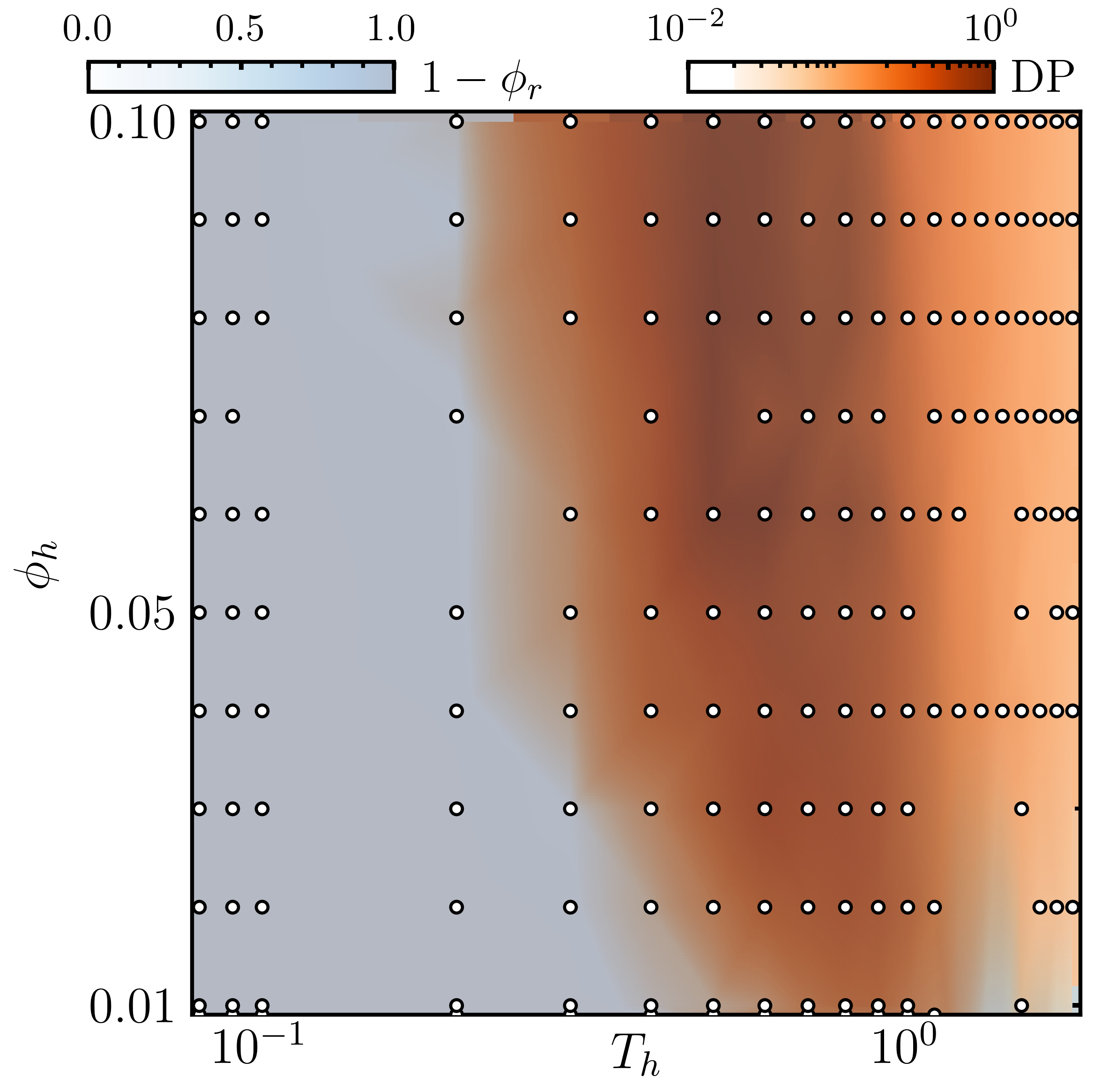}
    \caption{
Phase diagram of the active-passive mixture in the $(T_{h},\phi_{h})$ plane. 
The blue-white color scale indicates $1-\phi_r$, where $\phi_r$ is the fraction of rearranging particles (with blue denoting the arrested state and white the fluid state). The white-red scale indicates the demixing parameter $\mathrm{DP}$ (with red denoting a demixed state). }
    \label{fig:S7}
\end{figure}

We now explore a mixture of hot and cold dopants, characterized by the fraction of hot dopants $\phi_h$ and the hot-dopant temperature $T_h$. To quantify segregation between hot and cold particles, we compute the demixing parameter \cite{ManningPRL2024SI}. Because the mixture is compositionally unbalanced ($\phi_h<1/2$), we use the composition-corrected definition
\begin{equation}
\mathrm{DP}=\langle \mathrm{DP}_i\rangle
=\left\langle \frac{\frac{N_s}{N_t}-\phi_h}{1-\phi_h}\right\rangle,
\end{equation}
where $N_s$ is the number of homotypic neighbors of particle $i$ and $N_t$ is its total number of neighbors. With this normalization, $\mathrm{DP}=0$ corresponds to a well-mixed system, while $\mathrm{DP}=1$ indicates complete demixing.
Figure~\ref{fig:S7} shows the phase diagram in the $(T_h,\phi_h)$ plane. As in the active-passive mixture, sufficiently large $T_h$ fluidizes the system even at small dopant fractions. In contrast to active-passive mixtures, we do not observe a void-forming regime. Instead, at intermediate temperatures we observe a demixed phase, consistent with previous observations in 50/50 mixtures \citeS{FreyPRL2016SI,ManningPRL2024SI}.

\begin{figure}[h]
    \centering
\includegraphics[width=0.45\linewidth]{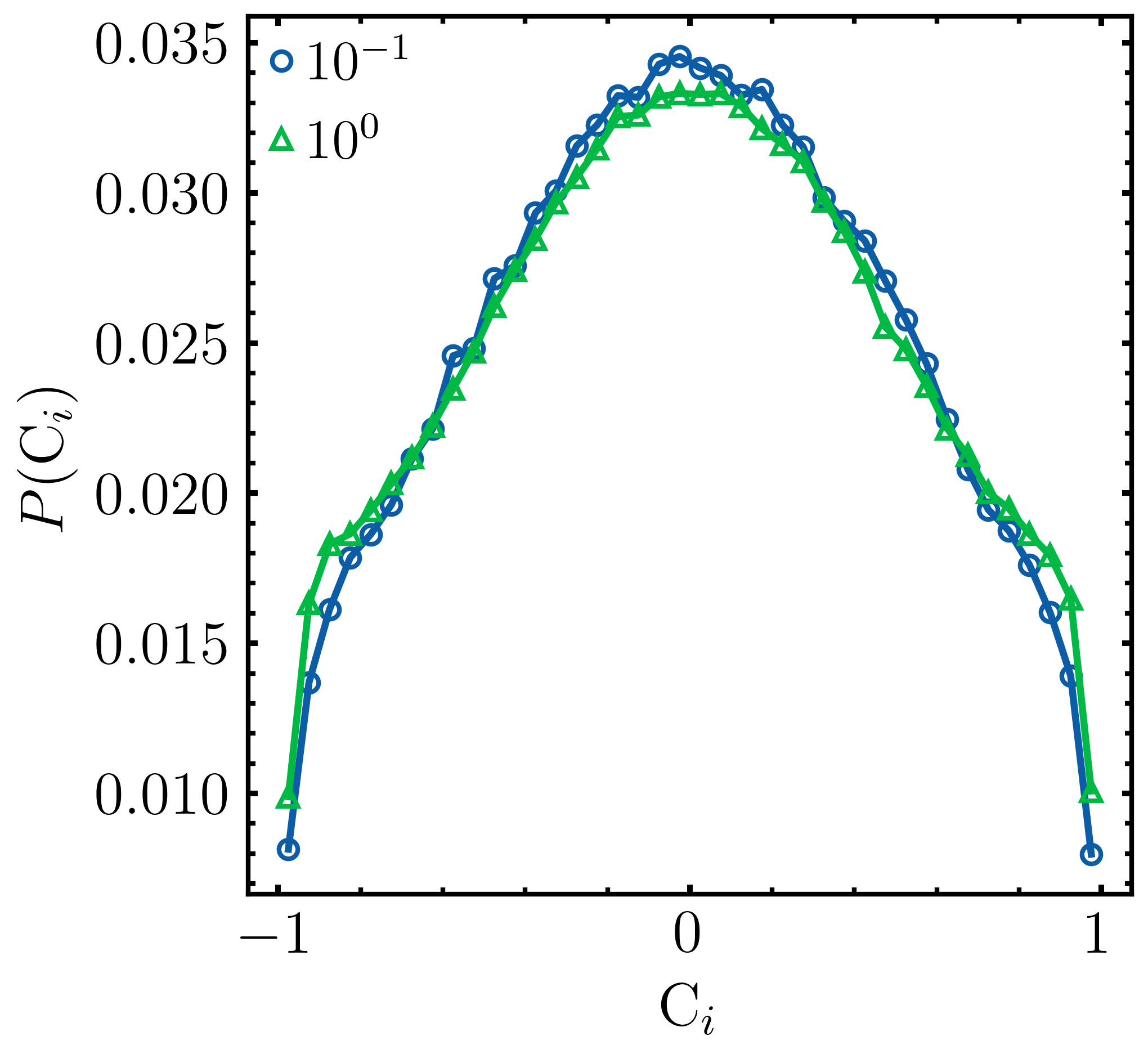}
    \caption{Probability distribution of cold-particle chirality, $P(\mathrm{C}_i)$, for $T_h=10^{-1}$ (blue circles) and $T_h=10^{0}$ (green triangles).}
    \label{fig:S8}
\end{figure}
Finally, we compute the probability distribution of the cold-particle chirality at two different temperatures. Figure~\ref{fig:S8} shows that, independent of temperature, the system does not exhibit chiral events; accordingly, the distribution is symmetric about zero.

\newpage
\section*{Supplemental References}

\end{document}